\documentclass{aa}
\usepackage{amsmath}

\usepackage{cleveref}
\usepackage{graphicx}
\usepackage{caption}
\usepackage{booktabs}
\usepackage{natbib}
\usepackage{placeins}
\bibpunct{(}{)}{;}{a}{}{,}

\newcommand{\CIIIHIIvCHIIIOH}{\ensuremath{\mathrm{C_3H_2/CH_3OH}}}
\newcommand{\HIICOvCHIIIOH}{\ensuremath{\mathrm{H_2CO/CH_3OH}}}
\newcommand{\HCOvCHIIIOH}{\ensuremath{\mathrm{HCO/CH_3OH}}}
\newcommand{\CNvHCN}{\ensuremath{\mathrm{CN/HCN}}}

\newcommand{\HCOPvHCN}{\ensuremath{\mathrm{HCO^+/HCN}}}
\newcommand{\HNCvHCN}{\ensuremath{\mathrm{HNC/HCN}}}
\newcommand{\CSvSO}{\ensuremath{\mathrm{CS/SO}}}
\newcommand{\SIOvSO}{\ensuremath{\mathrm{SiO/SO}}}
\newcommand{\CSvCN}{\ensuremath{\mathrm{CS/CN}}}
\newcommand{\CIIIHII}{\ensuremath{\mathrm{C_3H_2}}}
\newcommand{\CHIIIOH}{\ensuremath{\mathrm{CH_3OH}}}

\newcommand{\HCO}{\ensuremath{\mathrm{HCO}}}
\newcommand{\CN}{\ensuremath{\mathrm{CN}}}
\newcommand{\HCN}{\ensuremath{\mathrm{HCN}}}
\newcommand{\HCOP}{\ensuremath{\mathrm{HCO^+}}}
\newcommand{\HNC}{\ensuremath{\mathrm{HNC}}}

\newcommand{\SO}{\ensuremath{\mathrm{SO}}}
\newcommand{\SIO}{\ensuremath{\mathrm{SIO}}}

\newcommand{\density}[1]{\ensuremath{n_\mathrm{H}=10^{#1}\;\mathrm{cm}^{-3}}}

\newcommand{\deptemp}{\ensuremath{\phi_T}}
\newcommand{\depdens}{\ensuremath{\phi_{n_\mathrm{H}}}}

\begin{document}

\title{Understanding molecular ratios in the carbon and oxygen poor outer Milky Way with interpretable
machine learning.}
\titlerunning{Understanding ratios with interpretable ML}
\author{Gijs Vermariën \inst{1,2} , Serena Viti \inst{1, 3, 4}, Johannes Heyl \inst{4} \& Francesco Fontani \inst{5,6,7}}

\institute{
Leiden Observatory, Leiden University, P.O. Box 9513, 2300 RA Leiden, The Netherlands
\\ corresponding author email: vermarien@strw.leidenuniv.nl
\and
SURF, Amsterdam, The Netherlands
\and
Transdisciplinary Research Area (TRA) ‘Matter’/Argelander-Institut für Astronomie, University of Bonn, Bonn, Germany
\and
 Department of Physics and Astronomy, University College London, Gower Street, London, UK 
\and
INAF - Osservatorio Astrofisico di Arcetri,
              Largo E. Fermi 5,
              I-50125, Florence, Italy
\and
Max-Planck-Institut f\"{u}r extraterrestrische Physik, Giessenbachstra\ss e 1, 85748 Garching bei M\"{u}nchen, Germany
\and
LUX,Observatoire de Paris,PSL Research University,CNRS,Sorbonne
Universit\'e,F-92190 Meudon (France)
}

\abstract{
  The outer Milky Way has a lower metallicity than our solar neighbourhood, but still many molecules are detected in 
  the region. Molecular line ratios can serve as probes to better understand the chemistry and physics in these regions.
}{
  We use interpretable machine learning to study 9 different molecular ratios, helping us understand the forward connection
  between the physics of these environments and the carbon and oxygen chemistries.
}{Using a large grid of astrochemical models generated using UCLCHEM, we study the properties of molecular clouds
of low oxygen and carbon initial abundance. We first try to understand the line ratios using a classical analysis. 
We then move on to using interpretable machine learning, namely Shapley Additive Explanations (SHAP), to understand
the higher order dependencies of the ratios over the entire parameter grid. Lastly we use the Uniform Manifold Approximation and Projection technique (UMAP) as a reduction method to create intuitive 
groupings of models.
}{We find that the parameter space is well covered by the line ratios,
allowing us to investigate all input parameters. SHAP analysis shows that the temperature
and density are the most important features, but the carbon
and oxygen abundances are important in parts of the parameter space. Lastly,
we find that we can group different types of ratios using UMAP.
}{
We show the chosen ratios are mostly sensitive to changes in the carbon initial abundance, together with the temperature and density. Especially the 
\CNvHCN~ and \HNCvHCN~ ratio are shown to be sensitive to
the initial carbon abundance, making them excellent probes for this parameter. Out of the ratios, only \CSvSO~ 
shows a sensitivity to the oxygen abundance. 
}


\maketitle

\section{Introduction} \label{sec:intro}
In the outskirts of our Milky Way we find Giant Molecular Clouds (GMC) that play a crucial role in the process of star and planet formation.
Understanding the chemical composition within these clouds is essential to understand the star formation process in the Outer Galaxy (OG).
In these parts of the Galaxy, the environment is deprived of oxygen and carbon compared to our own solar neighborhood \citep{estebanRadialAbundanceGradient2017}. 
This lowered metallicity implies that there should be fewer atomic building blocks to build complex organic molecules (COMs).
These COMs are molecules with more than 6 constituent atoms, and are important
to understand the formation of prebiotic molecules  \citep{herbstComplexOrganicInterstellar2009}. 
Yet in recent observations, COMs are detected in several low metallicity environments, such as star forming regions in the outer galaxy \citep{shimonishiDetectionHotMolecular2021, bernalMethanolEdgeGalaxy2021} and the Magellanic Clouds \citep{
sewiloDetectionHotCores2018, sewiloDetectionDeuteratedWater2022,
shimonishiDetectionHotMolecular2023},  implying a chemical richness not expected in these environments.
Studying the chemical complexity as a function of metal poor gas is essential to better understand the star and planet formation in these low metallicity regions.
Even more recently, the project `Chemical complexity in the star-forming regions of the outer galaxy' (CHEMOUT) observed 35 of such star-forming cores at the 
edge of our own Milky Way, confirming their molecular richness \citep{fontaniCHEMOUTCHEMicalComplexity2022,fontaniCHEMOUTCHEMicalComplexity2022a,colziCHEMOUTCHEMicalComplexity2022, fontaniCHEMOUTCHEMicalComplexity2024}. 

GMCs typically have a low temperature (T$<$100K) and number densities of $n_\mathrm{H}>100\;\mathrm{cm}^{-3}$. The astrochemistry within these clouds
is driven by the fact that the gas is cold enough to `freeze' atoms and molecules onto the grains of the dust particles, allowing for ice chemistry to occur. Within these
ices, molecules can increase in complexity by reacting with one another, one of the main pathways being hydrogenation, and form larger molecules.
With modern telescopes, these molecules can be detected at an unprecedented rate, creating a compendium of many molecular observations of increasingly complexity.  
Computational models are then used to simulate the astrochemical processes that produce and destroy molecules within these clouds.
However, with a great variety of possible physical conditions, chemical histories and many different molecules, 
it becomes hard to interpret the observations. The uncertaintities on the elemental abundances of a region further complicates the process.  In this study 
we attempt to circumvent this issue by  investigating how several simulated molecular line ratios can be interpreted using machine learning.

In order to model the origin of the molecules in the outer regions of the Galaxy, we use the gas-grain chemical code UCLCHEM\citep{holdshipUCLCHEMGasgrainChemical2017} to model the regions as molecular clouds of constant density and temperature. 

We then use a combination of a classical analysis and interpretable machine learning to interpret a large grid of these models. 
Classical analysis of astrochemical models has been extensively used in the past to model and understand a variety of objects and astrophysical
processes \citep{bayetMolecularTracersHighMass2008, bayetMolecularTracersPdrDominated2009a, wakelamSensitivityAnalysesDense2010,
bayetHowCosmicRays2011, woodsFormationGlycolaldehydeDense2012a}.
These however could only cover a relatively small part of the parameter space, since both the generation and interpretation of a large number of models
was not yet feasible. With new machine learning methods \citep{haradaALCHEMIAtlasPrincipal2024} and especially interpretable machine learning methods \citep{heylStatisticalMachineLearning2023,heylUnderstandingMolecularAbundances2023,ramosFastNeuralEmulator2024a,grassiMappingSyntheticObservations2025},
the interpretation of large grids of models at once becomes feasible.
Interpretable machine learning  is a rapidly evolving field that concerns itself with providing insight into how machine learning models come to their prediction. The field
of interpretable machine learning is rapidly evolving as increasingly complex artificial intelligence methods require investigations as to why they work so well. However,
interpretable machine learning can also be used as a tool to help one understand nonlinear and complex classical processes.
We use the interpretable machine learning as a method to help us interpret our large parameter space, specifically with Shapley Additive exPlainers (SHAP) \citep{lundbergUnifiedApproachInterpreting2017}.
SHAP finds its origin in game theory \citep{shapleyAssignmentGameCore1971} and is especially useful for understanding the nonlinear forward connections between input and output, in this paper the physical parameters and ratios respectively.
The method quantifies the contribution of each of the input parameter to the output prediction, treating it as an additive game.
In order to better understand our astrochemical models, we train boosted regression forests and use the TreeSHAP algorithm to extract explainers for each of the ratios.

Since each model has six features and six corresponding SHAP contributions, the high dimensionality of the dataset remains. By plotting these together with a colormap, we only have three 
dimensions we can investigate at once. This however limits severely the
interpretability since it requires a quadratic number of plots to investigate the effects of each feature independently, and
often there are degeneracies in the plots. To alleviate this problem, we introduce the Uniform Manifold Approximation and Project Technique (UMAP)
\citep{mcinnesUMAPUniformManifold2020}
constructed using the SHAP contributions and the ratio itself. This method
results in a two-dimensional coordinate space, in which the models 
are arranged into a smooth manifold, grouping similar SHAP contribution vectors. By then using a colormap to represent the ratio, features and
SHAP contributions, we can investigate groups of ratios, their dependence
on the physical parameters and how SHAP clusters them together. This
allows us to investigate the SHAP values in the most informative two
dimensional representation, disentangling the degeneracies present
in classical two-dimensional plots.

In \Cref{sec:methods} we first describe the setup of the grid of astrochemical models and how we convert these to mock observations of molecular line ratios; then we
describe the theoretical framework of SHAP. In \Cref{sec:results} we start with a classical analysis of the mock observations, we then proceed to 
analyze the ratios using SHAP and UMAP. In \Cref{sec:conclusion} we conclude the paper.        
\section{Methods} \label{sec:methods}
Kinetic chemical codes have long been used to provide insight into the formation and destruction of molecules
in various astronomical contexts, e.g., those discussed in \cite{brownModelChemistryHot1988, millarGasPhaseReactions1991, vitiTimedependentEvaporationIcy1999a, ruaudGasGrainChemical2016a,rolligPDRCodeComparisonStudy2007}. 
These codes keep track of the total densities of various molecules as a function of time, space and/or visual extinction,
providing insight into their formation and destruction mechanisms, driven by both physics and chemistry. 
We specifically model the objects in the outer galaxy as dark clouds, without any energetic source (e.g. hot core or shock models),
consistent with the expected lower cosmic ray ionisation and radiation field in the outer galaxy.

\subsection{Modelling dark clouds with UCLCHEM}
The modeling of the chemical composition in dark clouds is done using the open-source gas-grain chemistry code UCLCHEM \citep{holdshipUCLCHEMGasgrainChemical2017}. 
This code allows us to model both the gas and grain chemistry in a time dependent manner. This  provides us with timeseries
that describe the abundances of each of the molecules present in the model. The modeling is done assuming istothermal
clouds of constant density. For these clouds we then vary 6 parameters, the number density, temperature, cosmic ray ionisation rate,
UV radiation field, initial elemental abundance of carbon and initial elemental abundance of oxygen. We explore cold molecular cloud models of several
densities ranging from $10^3\;\mathrm{cm}^{-3}$ to $10^7\;\mathrm{cm}^{-3}$ and temperatures of up to 100$\;$K. The 
cosmic ray ionisation rate goes from the typical galactic value \citep{indrioloH3DiffuseInterstellar2007} up to $10^3$ and the UV radiation field ranges
from 0.1 to 10 Habing. The elemental abundances of carbon and oxygen are depleted independently by up to a factor of 20
compared to solar values \citep{fontaniCHEMOUTCHEMicalComplexity2024, mendez-delgadoGradientsChemicalAbundances2022}. The range of all values and whether we sample them in a linear or logarithmic fashion
can be found in \Cref{tab:sobolgrid}.
The initial elemental abundances for the atoms that are not varied in the grid, can be found in \cref{app:elemental_abundances}.
The models are ran up to a time of $10^7$ years each; with a cloud radius of $R=0.5\mathrm{pc}$, which is consistent
with the lower limit of the regions in \cite{fontaniCHEMOUTCHEMicalComplexity2024}. This results
in clouds with visual extinctions of $A_V = 2.0$ at the lowest densities. This includes an edge visual extinction of 1 mag.
At the highest density, the visual extinctions reach up to $A_V\sim 10^4$.

\begin{table*}[h]
    \caption{The grid of parameters chosen for this study. The parameters are sampled using a Sobol sampling scheme.}
    \centering
    \begin{tabular}{|c|c|c|c|}
    \hline
    Parameter & Min & Max & Sample space \\
    \hline
    Density $n_\mathrm{H}\;(\mathrm{cm}^{-3})$ & $1 \times 10^3$ & $1 \times 10^7$ & log \\
    Temperature $T\;(\mathrm{K})$& 10 & 100 & linear \\
    Cosmic ray ionisation rate $\zeta\;(s^{-1})$& $1 \times 10^{-17}$ & $1 \times 10^{-14}$ & log \\
    Radiation field $F_{\mathrm{UV}}\;(\mathrm{Habing})$ & 0.1 & 100 & log \\
    Initial abundance of carbon $f_\mathrm{O}/f_\mathrm{O,\odot}\;(-)$  & $0.05 \times 1.77 \times 10^{-4}$ & $1.0 \times 1.77 \times 10^{-4}$ & linear \\
    Initial abundance of oxygen $f_\mathrm{C}/f_\mathrm{C,\odot}\;(-)$  & $0.05 \times 3.34 \times 10^{-4}$ & $1.0 \times 3.34 \times 10^{-4}$ & linear \\
    \hline
    \end{tabular}

    \label{tab:sobolgrid}
\end{table*}

In order to alleviate the curse of dimensionality of our 6-dimensional parameter space, we use Sobol sequence sampling \citep{sobolDistributionPointsCube1967}.
Often uniform random sequences are used to sample such spaces, but they have a high discrepancy. The 
high discrepancy can become a problem when trying to analyze the output of these non-linear models. Another
method would be to use grid or latin-hypercube sampling, which both guarantee that the marginal distribution for each of
the parameters is uniform, but grid sampling become intractable quickly and latin-hypercube sampling does not
guarantee a low discrepancy either. Sobol addresses both the computational and discrepancy shortcomings and allows
us to efficiently investigate the parameter space.
This results in a grid of $2^{16} = 65536$ models. Some of these models however do not
run successfully: this can be attributed to certain computationally stiff regimes where the freeze-out
onto the grains and desorption are in competition with each other, causing the timescales of the reactions
to become extremely short and expensive to solve. In this case UCLCHEM will choose to not integrate 
until the final time. We experimented with treating this missing data by both excluding the data points and substituting
the final value for the last-known value. The former method turns out to be the most effective since 
the latter tends to create spurious ratios that do not agree well with the distribution of neighbouring parameter sets.
Hence we exclude spurious data points.
With all the abundances simulated as a function of time, we can now compute the molecular line ratios. We choose to compute
the ratios at $10^5$ years. At this time, the gas phase species have not had a chance to fully freeze-out
onto the grains. This allows us to investigate a relatively young astrochemistry on the timescale
it takes for a young stellar object to form \cite{williamsIntroductoryLectureFrontiers1998}.  
In order to account for the fact that molecules with a low number abundance are not observable, we take the abundances for the model
and filter them based on a minimal abundance that is needed to result in an ``observable'' intensity. 
We take the lower limit of $x_i \ge 10^{-12}$, since this is conservatively the lowest abundance we can observe.
Any ratios with a non-detection in either its enumerator or denominator will thus not be taken into consideration. 
This provides the dataset of molecular ratios, which can help us the forward relationship between the physical
conditions and the chemical composition.

\subsection{Molecular ratios as tracers of physical conditions}
To probe the physical conditions and the chemical composition of various astrophysical regions, molecular line 
ratios serve as an essential diagnostics. On an extra-galactic scale, they have been used extensively to characterize
starburst galaxies \citep{haradaALCHEMIAtlasPrincipal2024,butterworthUnderstandingIfMolecular2022} and Active Galactic Nuclei \citep{konigMajorImpactMinor2018, useroMolecularGasChemistry2004,garcia-burilloMolecularGasChemistry2010}. 
On a galactic scale, ratios have been used to characterize molecular clouds \citep{penalozaCOLineRatios2018,tafallaCharacterizingLineEmission2021}.
We propose the use of three main groups of ratios, namely methanol based, hydrogen cyanide based and finally sulfur based ratios, many of which
have been used to probe different physical conditions and environments.
The ratio of \HIICOvCHIIIOH~ is connected to the formation of complex organic molecules in the ice phase and can also serve
as a probe of the formation timescales of massive star formation \citep{sabatiniEstablishingEvolutionaryTimescales2021}. 
When combined with cyclopropenylidene, the \CIIIHIIvCHIIIOH~ratio, has been employed to constrain the effects of the 
interstellar radiation field on  starless cores \citep{spezzanoDistributionMethanolCyclopropenylidene2020}
To further probe the formation processes of methanol, we can also combine it with its first hydrogenated
precursor, \HCO, providing insight into the formation pathways \citep{bacmannOriginGasphaseHCO2016}.
Physical conditions can also be probed by ratios such as \HNCvHCN. This ratio is currently being debated to be a strong predictor
of the radiation field \citep{haradaTemperatureFUVTracer2024a}, cosmic ray ionisation \citep{behrensTracingInterstellarHeating2022} or temperature \citep{hacarHCNtoHNCIntensityRatio2020}.
Another \HCN~based ratio is \HCOPvHCN, which can trace energetic environments such as AGNs \citep{butterworthUnderstandingIfMolecular2022}.
We also include the \CNvHCN~ ratio, as it is sensitive to both carbon and oxygen abundances \citep{milam1213Isotope2005}, 
serving as an effective tracer of dense gas \citep{wilsonNearlyConstantCN2023} and is associated with evolved starbursts \citep{haradaALCHEMIAtlasPrincipal2024}.
The first sulfur based ratio we investigate is \CSvSO, which can probe the oxygen to carbon ratio directly 
in protoplanetary disks 
\citep{semenovChemistryDisksXI2018,galMoleculesALMAPlanetforming2021}
and provide
a chemical clock for massive star formation \citep{liSulfurbearingMoleculesMassive2015}.
The molecular ratio \SIOvSO~can be used to infer physical parameters of energetic systems (such as shocks) with grain processing present \citep{jamesRevealingPhysicalConditions2021a,codellaMolecularOutflowsIntermediatemass1999}.
Lastly, \CSvCN~can be used as a dense gas tracer \citep{wangCN21CS2022}.
We then proceed with analyzing these ratios using interpretable machine learning.

\subsection{Interpretable machine learning with SHAP}
Modeling the chemistry of astronomical objects with grids of chemical models data is one of the classical methods to understand the forward connection between physical and chemical parameters and molecular abundances and ratios.
In the past, the application of computational models to astrochemistry was constrained by the computational cost of running them for different parameter configurations. Nowadays, computational resources are great enough that we can generate large volumes of simulations.
This introduces the problem that high-dimensional simulation grids with
several output ratios become increasingly hard to interpret by hand. Historically, conditional and marginal plots have been the go-to method to interpret parameter studies,
but these disregard higher-order interactions and require extensive expert knowledge to interpret. Interpretable machine learning addresses this by providing
model-agnostic methods to understand non-linear models. Model-agnostic interpretation methods can be easily cast into a sampling, intervention, prediction, aggregation
framework (SIPA) \citep{scholbeckSamplingInterventionPrediction2020} and distinguished into two broad subcategories, global and local methods.
Global methods are akin to the methods that astrochemists have been using in astrochemical literature extensively, namely partial dependence plots and even Principal Component Analysis surrogates more recently.
Thus we focus on the usage of local methods, as they can provide more insight in sub-regions of the dataset by explaining the individual examples. This
has been shown to be a powerful tool both in astronomy \citep{heylStatisticalMachineLearning2023,heylUnderstandingMolecularAbundances2023,ramosFastNeuralEmulator2024a,grassiMappingSyntheticObservations2025}, but also in fields like geophysics and biomedicine. Two popular
global methods at this time are Local interpretable model-agnostic explanation (LIME)\citep{ribeiroWhyShouldTrust2016} and SHAP \citep{lundbergUnifiedApproachInterpreting2017}; the former tries to construct local surrogates 
for each individual prediction whereas the latter tries to provide explanations by using a global interpretation method. In this work we choose 
SHAP because we are interested in the behavior of the ratios over the whole physical range and its subsets, not in a sensitivity study of individual samples. 

SHAP is an efficient approximation of the game-theory concept of Shapley values. These Shapley values are a method to evaluate how much a feature contributes
to the output of a model by considering all possible player coalitions and the cost of each of these. An illustrative example is sharing a cab that brings home several individuals, where each addition or removal of a person to the coalition results in a different cost \citep{molnarInterpretableMachineLearning2022}. 
This can be formalized into a Shapley value $\phi_j$ for 
each feature $j$. These values satisfy the following properties:
\begin{itemize}
    \item Efficiency - all feature contributions together must sum to the output minus the expected value of the model.
    \item Symmetry - if two features contribute equally across all coalitions, their SHAP value is identical
    \item Dummy -  if a feature does not change the output, its Shapley value is zero
    \item Additivity - if you add two games together by summing the outputs, the SHAP values are the sum of the individual game's Shapley values.
\end{itemize}
Unfortunately, the explicit computation of the Shapley values can quickly become prohibitively expensive as all coalitions must be evaluated
in order to obtain the exact value. SHAP addresses this issue by instead computing the contribution of each feature as the
weighted average of the marginal contributions. This shows us that each prediction must be a sum of the feature
explanations plus the expected value of the predictor:
\begin{equation}
    \hat{g}(x) = \sum_j \phi_j + \mathbb{E}(g(x)).
\end{equation}
The Shapley value $\phi_j$, also referred to as impact or contribution, 
determines how much each feature (e.g. density) has contributed to one realization
(e.g. \SIOvSO) according
to the model. Concretely, we use the contributions to
quantify the impact of each invdividual feature on each ratio sample.

This marginal contribution for each feature can be approximated in several ways such as 
a linear explanation model with kernelSHAP \citep{lundbergUnifiedApproachInterpreting2017}, 
a neural network with deepSHAP \citep{chenExplainingModelsPropagating2019} and a forest model treeSHAP \citep{lundbergLocalExplanationsGlobal2020}. We will use the latter, since it 
provides a computational effective method that is particularly well suited for the pre-generated tabular data on a grid but instances of deepSHAP have also been used in astronomy \citep{ramosFastNeuralEmulator2024a,grassiMappingSyntheticObservations2025}.
As the name suggests, treeSHAP relies on decision trees, for this specific use case we will use boosted regression forests. Regression forests are
a combination of decision trees that utilize continuous data, with boosting referring  to the concept of sequentially combining weak predicting trees
that become a strong predictor when combined into an ensemble \citep{hastieElementsStatisticalLearning2009}. The structure of the tree, descending down a path of decisions, causes the
number of possible coalitions to be constrained, alleviating the computational complexity of computing the SHAP values. 

\subsection{Uniform Manifold Approximation and Project technique (UMAP)}
In order to reduce datasets to lower dimensions, the Uniform Manifold Approximation and Projection technique was developed \citep{mcinnesUMAPUniformManifold2020}.
Similar in nature to PCA \citep{pearsonLIIILinesPlanes1901} and t-SNE \citep{vandermaatenVisualizingHighDimensionalData2008}, it allows one to create lower dimensional representation of high dimensional data
that can aid in clustering, interpretation and feature importance. The method assumes the data points lie on a Riemannian manifold within a 
high-dimensional space and tries to find a mapping between the two that preserves both the local and global structure. This resulting
low-dimensional manifolds have been shown to help greatly in the classification in several astronomical contexts such as Auroral Dynamics \citep{lambCorrelationAuroralDynamics2019},
fast radio bursts \citep{chenUncloakingHiddenRepeating2022} and low metallicity stars \citep{kaneHuntGalacticFossils2023}. 

The algorithm tries to construct a weighted graph in high-dimensional space, connecting neighbors together. By constructing the K nearest neighbor (KNN) graph,
it captures the local structure of the data. They then construct a mapping function between the high-dimensional space onto the lower-dimensional space,
trying to preserve the nearest neighbors with a cross-entropy loss function.

Since our dataset is sampled on a regular grid, and we only have the ratio as a meaningful feature, the KNN algorithm will not perform well at the task of trying
to construct a meaningful representation. If we, however, interpret the impact of each feature as a component of a vector, one SHAP explanation can be seen as
a 6-dimensional vector whose values add up to the ratio. We use these 6 SHAP features, together with the ratio, as the input into the UMAP algorithm. 
The most important hyperparameters are the number of neighbors $k$, the minimum distance between points, the low dimensional representation $d_{min}$,
and the weighting of the loss of the SHAP values versus the loss of the ratios themselves $w_{ratio}$. After manually tuning these, we choose $k=100$,
$d_{min}=\{0.1,0.5\}$ and a $w_{ratio}=0.1$. The goal 
of this tuning was to obtain smooth manifold that was not too compact but still clustered, allowing the manifold to both highlight different regions and changes as a function of the parameters; the chosen set of parameters reflects a good tradeoff between these two extremes.
The loss for the SHAP features is computed using the cosine distance, whilst the loss for the
ratio is computed using euclidean distance. This reflects the fact that we treat the SHAP contributions as a vector, whereas the ratio is added as measure to break degeneracies in the aforementioned vector space.

\subsection{Interpreting molecular line ratios with SHAP}
As described in Section 2.1, we start by using UCLCHEM to simulate each of the models on the parameter grid. This
then results in a timeseries for each molecule. From this timeseries we obtain the value at $10^5$ years.
We then apply the observational threshold for each molecule, if either the denominator and enumerator exceed the observational threshold, we compute its log-value and add it to the dataset.
With a dataset for each ratio, we split it into a training set and a test set, containing 70\% and 30\% of the set
respectively.
The train dataset is used to train a regression forest, with the physical parameters as input and the molecular line ratio as output. The hyperparameters of the regression forest are optimized using the Optuna framework \citep{akibaOptunaNextgenerationHyperparameter2019} with the test error as the optimisation target. 
We take the optimal configuration, and train the regression forest once more to generate the 
final predictor model. The SHAP values are then computed using this regression forest and the `TreeSHAP' algorithm as implemented in the SHAP package \citep{lundbergUnifiedApproachInterpreting2017}. In the end, this gives us the
SHAP values for each sample in the dataset. This is then repeated for each molecular line ratio, resulting in 9 distinct regression forests and SHAP explainers.
Finally, we feed these SHAP values and the ratios into the UMAP algorithm. This gives us a two-dimensional embedding for the 
data and help us interpret both the SHAP values, input features and ratios.

\section{Results} \label{sec:results}
In order to better understand the SHAP contributions, we first 
analyze the results in line with a classical sensitivity study of the ratios and their dependence on temperature and density. 
We then look at the relative importance of each input feature,
investigating the order of importance and the nonlinearity
of its impact.
Lastly, we investigate further the impact by plotting the
ratios, feature values and impacts on a low dimensional manifold,
grouping together similar models and revealing nonlinearities
in the ratios.
\subsection{The molecular ratios: from chemical modeling to ``observable'' abundances}

\begin{figure*}
\includegraphics[width=\linewidth]{"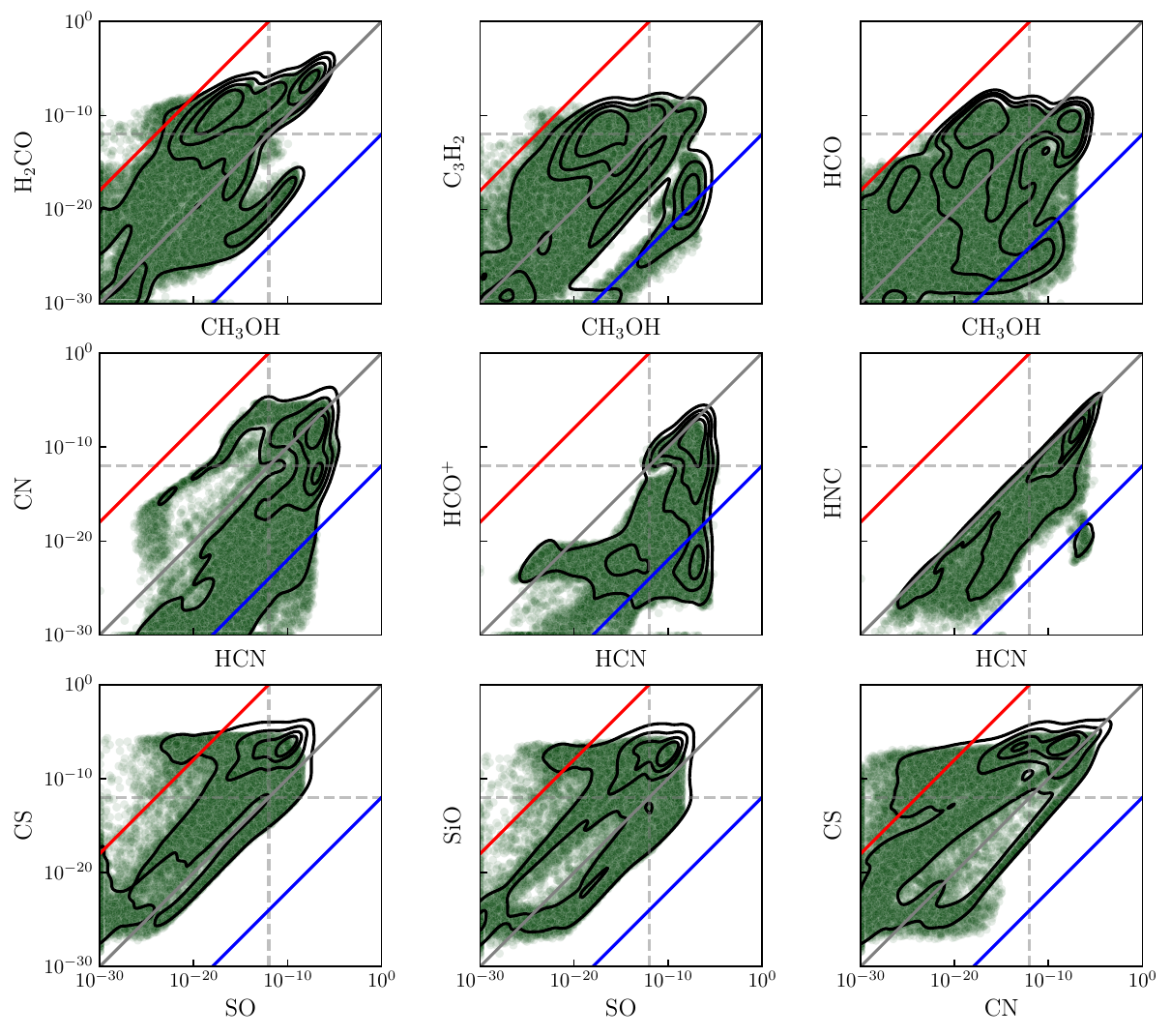"}
\caption{Plots showing fractional abundances for each of the ratios discussed in this paper. Contour levels of a kernel density estimate are added to highlight the distribution, each representing 20\% of the distribution. For both we show the
"observational" limit of $10^{-12}$ that is used throughout the paper. Only the ratios above either observational threshold are used for 
training the SHAP explainers. The blue line represents all log-ratios of $-12$,
the red line represents the log-ratios of $12$.}
\label{fig:results:abundance-pairplots}
\end{figure*}

Before reducing the fractional abundances to a ratio, we first inspect the distribution of the two species with respect
to each other. A plot of the two constituents of each ratio
is displayed in \Cref{fig:results:abundance-pairplots}.
We distinguish between four scenarios: only A is detected, both A and B are detected, only B is detected and lastly
neither are detected. This is represented by the 4 regions in the figures, with the 
dashed lines representing the observational limits. It can be seen that for every one of the ratios, part of the distribution lies in the range where both molecules can be detected. 

\begin{figure}
\centering
\includegraphics[width=1.0\linewidth]{"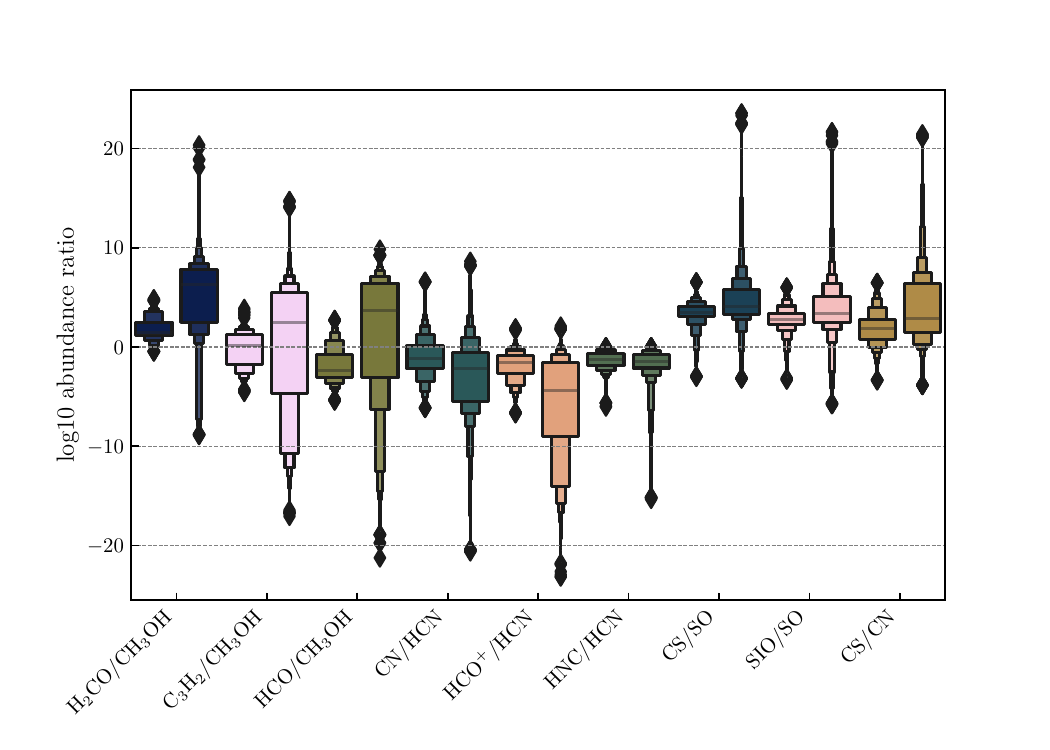"}
\caption{The distribution of each of the ratios that exceed the detection limit for both molecules. The left distribution for each ratio
is where both molecules are detectable whilst the right distribution for each molecule is where either molecule is detectable. The line in the central box is the median, the central box contains 50\% of the ratios, the 
next upper and lower boxes together contain 25\% and so forth, with outliers plotted as diamonds.}
\label{fig:results:ratio-hist}
\end{figure}

In order to extract the most information out of the grid of models, we choose to include any sample that detects either molecule. Even though the
more extreme ratios in the region where only one molecule is detected cannot be observed directly, they can still be useful when combined 
if an upper limit can be derived from the observation. 
The distribution of all ratios can be found in \Cref{fig:results:ratio-hist}.
The statistics for the ratios filtered by either detections can be found in \Cref{tab:results:distribution}. In the rest of the analysis, the
data will always be filtered for a detection in either molecule.

\begin{table*}
\caption{Statistics of the distribution of ratios with either molecule detected shown in \Cref{fig:results:ratio-hist}.}\label{tab:results:distribution}
\begin{tabular}{lrrrrrrr}
\toprule
 & count & mean $\mu$ & std $\sigma$ & min & median & max & $\sum_j{|\phi_j|}$\\
\midrule
$\log_{10}$(H$_2$CO/CH$_3$OH) & 40091 & 5.06 & 3.89 & -8.90 & 6.27 & 20.34 & 3.94 \\
$\log_{10}$(C$_3$H$_2$/CH$_3$OH) & 27418 & -0.01 & 6.81 & -17.03 & 2.44 & 14.74 & 7.14 \\
$\log_{10}$(HCO/CH$_3$OH) & 28873 & 1.12 & 6.51 & -21.23 & 3.64 & 9.84 & 7.27 \\
$\log_{10}$(CN/HCN) & 54167 & -2.90 & 3.81 & -20.62 & -2.13 & 8.62 & 4.77 \\
$\log_{10}$(HCO$^+$/HCN) & 53076 & -5.87 & 5.25 & -23.14 & -4.40 & 2.09 & 5.88 \\
$\log_{10}$(HNC/HCN) & 53033 & -1.78 & 1.91 & -15.33 & -1.42 & 0.00 & 1.77 \\
$\log_{10}$(CS/SO) & 56002 & 4.56 & 2.60 & -3.21 & 4.04 & 23.54 & 3.11 \\
$\log_{10}$(SiO/SO) & 55397 & 3.72 & 2.76 & -5.79 & 3.36 & 21.67 & 3.00 \\
$\log_{10}$(CS/CN) & 55692 & 3.93 & 3.59 & -3.86 & 2.89 & 21.46 & 4.63 \\
\bottomrule
\end{tabular}
\tablefoot{The first column denotes the number of models
that satisfy the detection conditions, out of a total of 64949 models. The
middle columns describe basic statistics of these ratios. The last column
describes the sum of the average absolute importance for each ratio.}
\end{table*}

In order to visualize the dependence of the ratios on the temperature and density, we  plot
the ratios as a colormap, as can be seen in \Cref{fig:results:ratio-density-temperature}. A plot
with both the denominator and enumerator of the ratios' fractional abundance can be found in \Cref{app:enum_denom}.

\subsubsection{Methanol based ratios}
Starting with the upper left ratio, \HIICOvCHIIIOH, there are two general distributions in density-temperature space:  
a low-density distribution with densities up to $10^6\;\mathrm{cm}^{-3}$,
and a high-temperature-density distribution with densities higher than $10^{6.2}\;\mathrm{cm}^{-3}$ and temperatures starting at 87$\;$K.

The low-density distribution is split into two parts, with a divide at $~$ 30$\;$ K. Below this divide, the formation
of both methanol and formaldehyde happen quickly and at a positive ratio, and at $t=10^5$ years both species peak in the gas
phase and then quickly freeze out completely onto the grains. 
Above this divide, the formation pathway of methanol on the grain becomes much less efficient and less methanol can be
desorbed into the gas phase, whilst the formation of formaldehyde is not affected as much.
This results in a positive ratio since methanol is barely above the detection threshold whilst formaldehyde achieves
fractional abundances of up to $10^{-8}$.

The general pattern of the \CIIIHIIvCHIIIOH~ratio is similar, but it lacks the distribution with a very negative log-ratio below 30$\;$K and a clear gap between the low and high densities. The negative distribution below the threshold in the low-density regime is due 
to the less effective formation of \CIIIHII~at these lower temperatures, whilst methanol is still peaking. This distribution is intersected by the blue line in \Cref{fig:results:abundance-pairplots}. 
Above this border the formation of \CIIIHII~increases, resulting in positive ratios. For the high-density-temperature
distribution the \CIIIHII~is already decreasing, whilst the \CHIIIOH~is peaking, resulting in a negative ratio.

The last methanol based ratio, \HCOvCHIIIOH, has a negative high-temperature-density distribution. Again, in the main distribution, a divide at 30$\;$K is present, with a negative horizontal gradient between temperatures
of 30 and 45 $\;$K. Between these temperatures, as density increases, the methanol becomes more abundant with  \HCO~being relatively constant in abundance. At densities well above \density{5}, the \HCO~starts freezing out, whilst the methanol is still more abundant, resulting in a small region with negative ratios at low temperatures.
Below 30$\;$K, the methanol becomes very abundant in the gas phase. This results in an even more negative ratio
for this region. Above 45 $\;$K, the formation of methanol is again less efficient, whilst \HCO~is more abundant and depleting
at a slower rate, resulting in positive ratios. The high-density-temperature distribution
has only detections of methanol, resulting in very negative distributions. This is also reflected by the distribution that intersects the blue line in \Cref{fig:results:abundance-pairplots} 

\subsubsection{\HCN~based ratios}
The hydrogen cyanide ratios are detectable almost everywhere with the exception of the high density, low temperature part of the parameter space. 

The leftmost part of the distribution is dominated by the photodissociation of \HCN~into \CN, resulting in a positive ratio. As the density increases,
the ratio starts to tend towards being dominated by \HCN. 
In the lower right area, at temperatures below 40$\;$K and densities above \density{5}, the chemistry
is dominated by a fast freezing out of \CN~, whilst \HCN~ freezes out on a slower timescale.
For the \HCOPvHCN~ratio, a similar pattern emerges, but now without positive ratios on the right part of the distribution.
At lower densities, the ratio is close to unity, but then as the density increases, there is less photochemistry and the
log-ratio becomes increasingly negative.
The isomer ratio \HNCvHCN, only has negative log-ratios, driven by an effective isomerisation pathway at higher temperatures: $\mathrm{H}+\mathrm{HNC}\rightarrow\mathrm{HCN}+\mathrm{H}$
~\citep{hacarHCNtoHNCIntensityRatio2020}.
The log ratio is the closest to being zero in the high temperature and very low temperature regime with densities between
$10^{4}$ and $10^{6.5}\;\mathrm{cm}^{-3}$. In the other regions, the \HCN~ strongly dominates over the \HNC~.

\subsubsection{Sulfur based ratios}
The sulfur based ratios trace out a similar region of the parameter space. The \CSvSO~log-ratio is positive in most of the 
parameter space, only at the lowest temperatures of the densities above $n_\mathrm{H} > 10^{5}\;\mathrm{cm}^{-3}$ the 
ratio becomes negative. This local negative ratio happens as the CS starts freezing out, whilst the SO does not. 
The \SIOvSO~ratio and the  \CSvCN~ show a similar pattern in the parameter space.
However, for the latter,  along the diagonal of the upper right of the parameter space, there is an 
interesting split. Below the split, the log-ratio is zero, as both CS and CN reach high abundances. Above the split,
the CN is depleted, causing the log-ratio to become positive.

This exploration of the distribution of the ratios in temperature-density space provides the context for the explainable machine learning 
methods applied to it. We can now investigate the order of importance of the physical parameters to better understand what
influences the ratios most.

\begin{figure*}
    \centering
    \includegraphics[width=\linewidth]{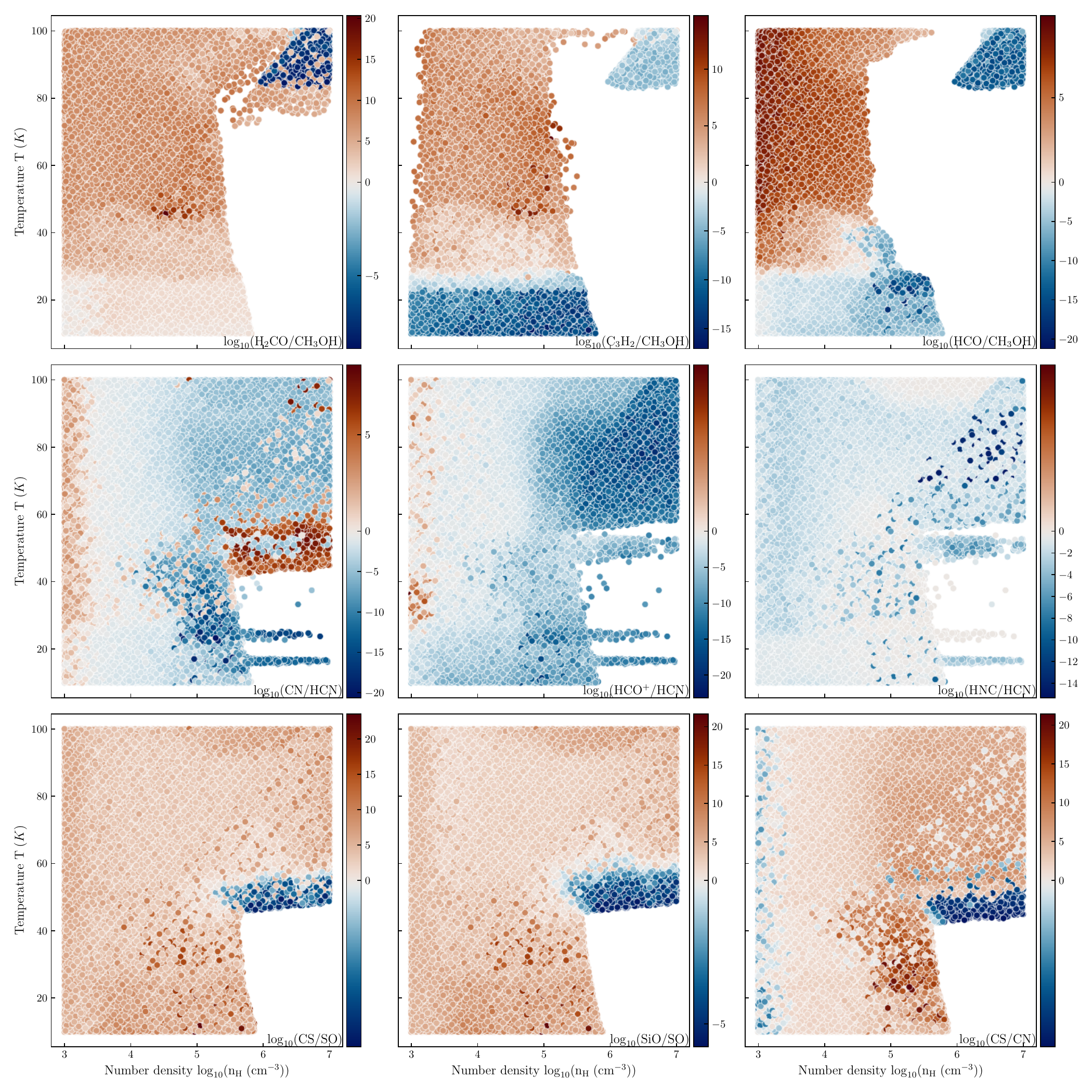}
    \caption{The distribution of the log of the ratios as a function of both density and temperature.}
    \label{fig:results:ratio-density-temperature}
\end{figure*}

\subsection{Mean SHAP impact -  ranking the physical features}
\begin{figure*}
    \centering
    \includegraphics[width=\linewidth]{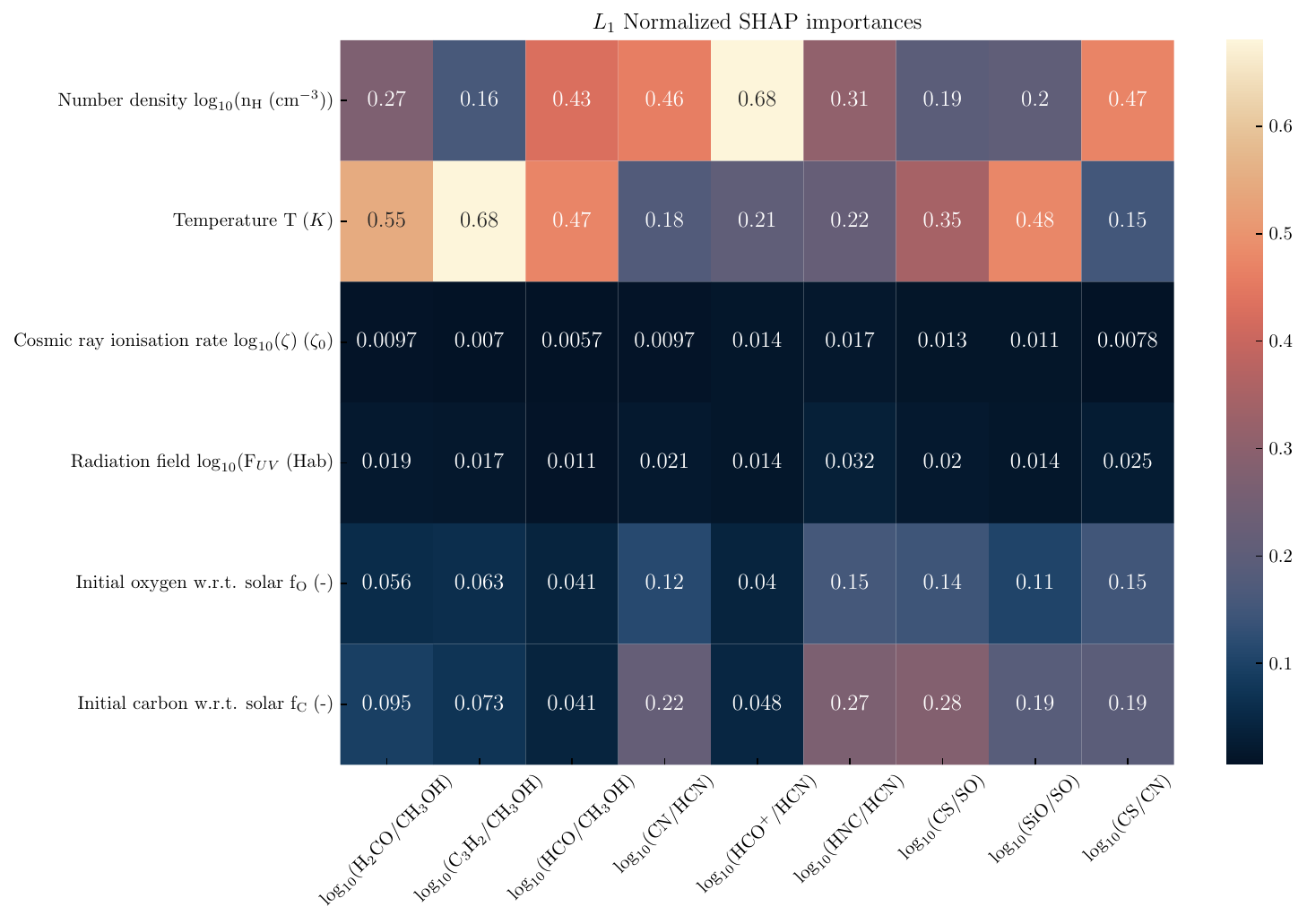}
    \caption{The relative importances of each physical parameter for all of the ratios.}
    \label{fig:results:shap-importances}
\end{figure*}

For each of the models, we compute the normalized feature importances, which is defined as the individual contribution
divided by the sum of the contributions for each ratio. This
explains the relative importance of each feature per ratio. 
We plot its values for each ratio in the heatmap in \Cref{fig:results:shap-importances}. 
The 
importances confirm that indeed the temperature and density are almost always the most important features, with the oxygen and carbon abundances sometimes close contenders. 
For example, for \CSvSO, the carbon abundance is more important than the number density, which is in line with the temperature-density distribution
in \cref{fig:results:ratio-density-temperature}, where the gradients of the ratios are small in the direction of the density. 
This order of features is useful for
distinguishing how sensitive features are, but it does not capture any information about how the ratio
is impacted exactly as a function of each feature, which is where SHAP summary plots come in.

\subsection{SHAP summary plots - interpreting first order interactions.}
\begin{figure*}
\includegraphics[width=\linewidth]{"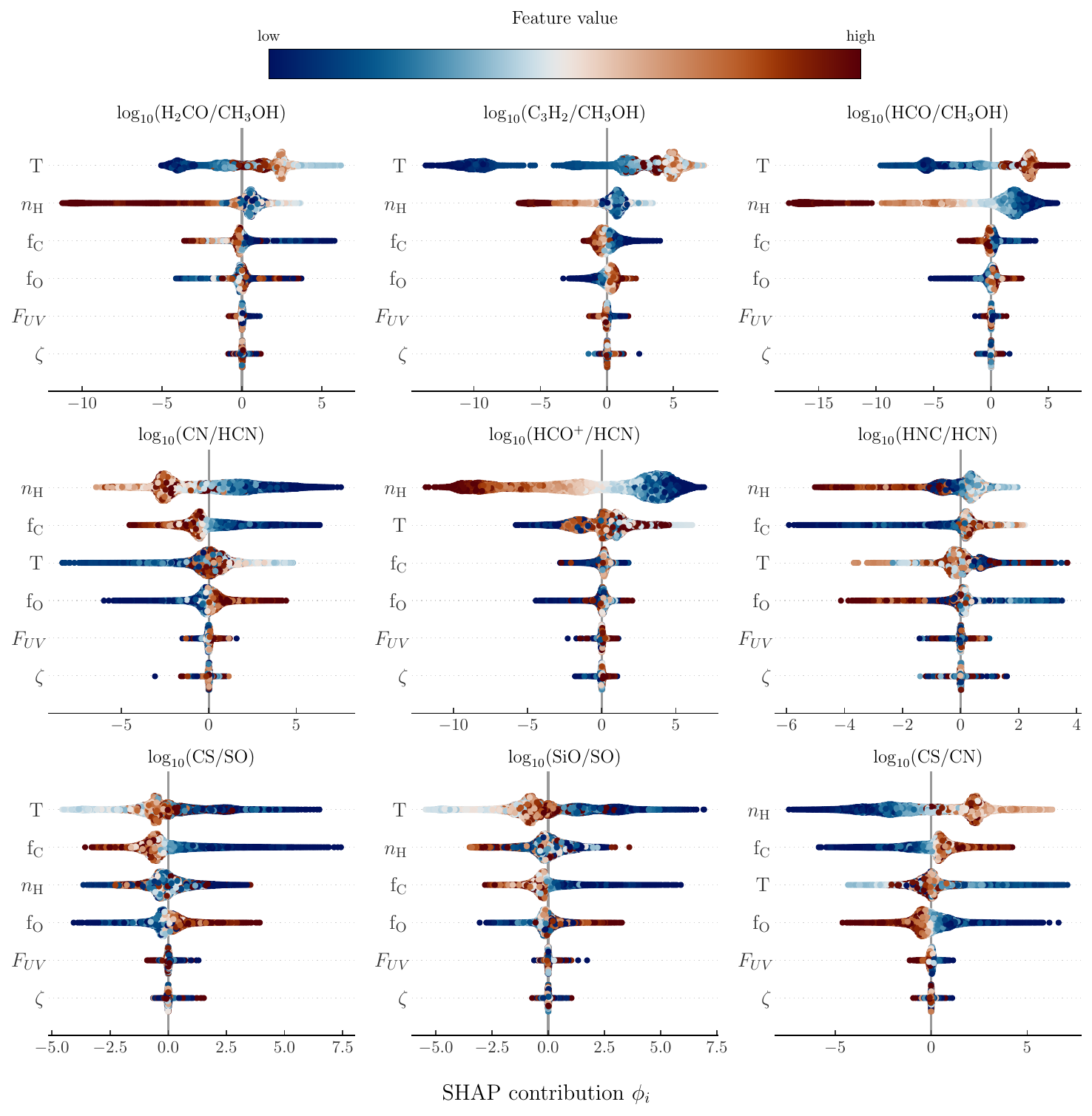"}
\caption{All of the SHAP values for each of the ratios. The features are sorted by mean absolute impact as shown in \Cref{fig:results:shap-importances}. 
Every colormap is normalized to the detectable range for each individual feature per ratio.}
\label{fig:results:shap_summaries}
\end{figure*}

The SHAP summary plots, as can be found in \Cref{fig:results:shap_summaries}, show a scatter point for each feature for each of the samples. Each sample
is then represented by six points across the parallel feature axes, with each impact color coded
with its respective value.
A positive SHAP contribution means the 
log-ratio was increased by the feature value and vice versa.

For all the methanol based ratios, the most important feature is temperature. The contribution of the temperature \deptemp is
monotonously increasing with its value for all ratios. Second is the impact of the
density \depdens , which is inversely proportional to the density. The impact
of carbon and oxygen are the third (inversely proportional to the feature value) and fourth (proportional to the feature value) most important features, respectively.

For the \HCN~based ratios, the density is the the dominant feature, with its impact
inversely proportional to the density. For the first and third \HCN~ based ratios,
\CNvHCN~and \HNCvHCN, the carbon is the second most important feature, with temperature coming
in third. For \HCOPvHCN, the temperature is the second most important feature, and
its shows no monotonous increase or decrease in impact. 
Interestingly, for \HNCvHCN, the temperature is not its most important feature, in contradiction with what has been found before, at least at low temperatures \citep{hacarHCNtoHNCIntensityRatio2020} as well as the feature importance findings of \cite{heylUnderstandingMolecularAbundances2023},
where temperature had the largest impact; these models used two modeling stages, in combination with higher densities and temperatures whereas our models assume static clouds. However still, the total metallicity in \cite{heylUnderstandingMolecularAbundances2023} and initial carbon abundance in this study both being the second most important features is consistent between the two SHAP explainers.
\CSvSO~is a ratio that has a strong dependence on temperature with the carbon abundance following closely. 
For \SIOvSO~ratio the
density is now the second most important feature, but the distribution of its impact
is complex.
The last ratio, \CSvCN, shows a strong impact for density, with the minimum and maximum densities having
negative contributions, whilst medium densities have a positive impact. 
The carbon dependence
show a clear proportional relationship. The temperature is inversely dependent,
with a medium temperature distribution with negative impact. 
Lastly, the oxygen
dependence shows a clear inverse proportional dependence.

The individual impact of each feature, while  providing insight into the ratios, does not tell us 
how two features depend on each other. In order to reveal higher order dependencies, we thus
rely on dependence plots, which show the dependence of one impact on both the feature itself
and others.

\subsection{UMAP plots - interpreting higher order interactions.}
The main benefit of using a statistical predictor and computing its SHAP values is that we can now
interpret the dependencies between the different features and
the ratios themselves. The UMAP manifold is generated using
the SHAP impacts and the ratios themselves, creating a 
convenient two dimensional representation of the data,
where both local and global features are relevant. This
allows us to identify distinct groups of models, that behave
similarly, and see how the features influence the ratio
within these groups.

We now proceed to analyze the ratios projected onto a 
two-dimensional manifold computed using the SHAP values and ratios, this way we can investigate the
clustering of similar models and see how they are dependent on 
input features.

\subsubsection{\HIICOvCHIIIOH}
\begin{figure*}
    \centering
    \includegraphics[width=\linewidth]{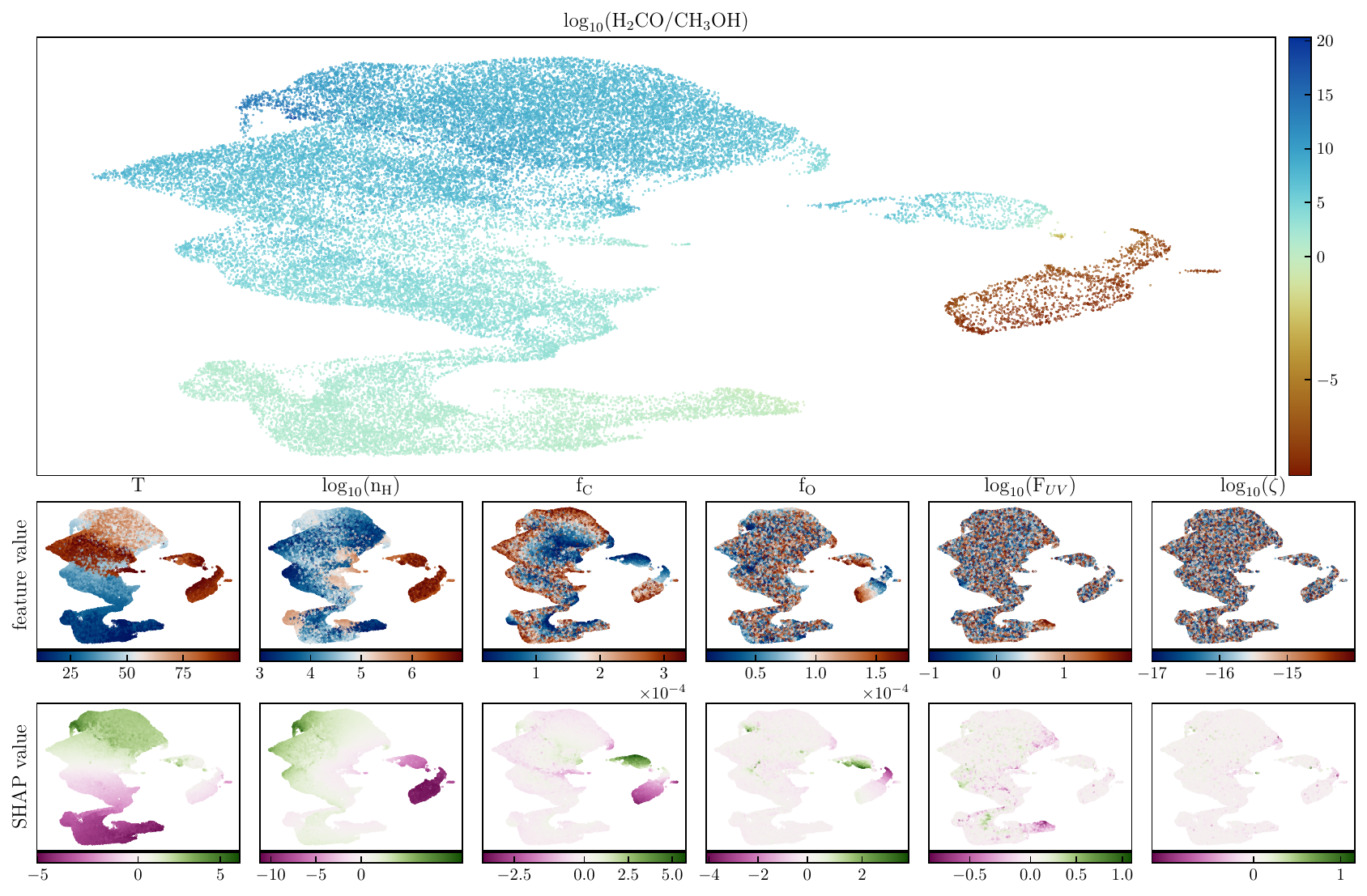}
    \caption{The \HIICOvCHIIIOH~ratio, feature and SHAP values plotted on 2-dimensional manifold using the ratio and the SHAP values.
    The manifold consists of a broad left region with two separate right lobes.} 
    \label{fig:results:h2coch3oh:umap}
\end{figure*}

For the \HIICOvCHIIIOH ratio, the UMAP plot 
can be seen in \Cref{fig:results:h2coch3oh:umap}. This plot shows that we can  
cluster the high temperature and low density models, on the top left side of
the manifold, with a positive impact of both features. This results in very positive ratios.

If we then move to the bottom
of the manifold, the temperature is low and the density is low as well; in this
case the contribution of the temperature is negative and the contribution of the density
ranges from small to negative. Especially interesting are the two lobes in the
right part of the plot. These regions correspond to the hidden distribution 
as can be seen in top right of \Cref{fig:results:ratio-density-temperature}. However,
this time we can clearly distinguish that, for the middle distribution, the 
impact of the temperature is positive, with a slightly negative density impact,
low carbon abundance, but a very positive carbon impact, high oxygen abundance and also
a positive oxygen impact. The rightmost lobe on the other hand has a high
carbon abundance distribution, with neutral temperature impacts, very negative
density impacts, high carbon abundances with negative impact and lastly a gradient
in the oxygen abundance, with a impact going from very negative to positive. This is all related to the distribution of methanol dominated ratios.

\graphicspath{{figures/time_1e5/umap}}

\subsubsection{\CIIIHIIvCHIIIOH}
We investigate the UMAP manifold, displayed in \Cref{fig:results:c3h2ch3oh:umap},
which clusters the distribution into three regions. The left region contains the lowest temperature gas,
having a negative temperature contribution. We can see that as the density increases along
the vertical axis, the SHAP contribution for the temperature decreases. In this region the carbon
impact is negative, with no clear pattern in its value. The oxygen value however shows a clear 
radial gradient, connected with a negative SHAP contribution in the center of the region and a positive
one on the outskirts. 

The right, more extended region, contains the medium temperature, low density gas, with positive contributions.
The carbon abundance shows a clear gradient from positive to negative as we move downwards, with the SHAP contribution
going from positive to negative. The pattern for the oxygen shows a radial pattern with the inner region having 
negative contributions with low values, and the outer region having positive contributions. 
Lastly there is a lobe on the bottom: this is the high temperature, high density gas, with a positive
temperature impact and a negative density impact. Here we can again see that the contributions of carbon
and oxygen influence the ratio.

\subsubsection{\HCOvCHIIIOH}

The manifold for this ratio is displayed in \Cref{fig:results:hcoh3oh:umap}, it shows an elongated
distribution of the gas. The continuous part of the manifold contains the distribution
of warm gas on the right side, with low to medium densities. This gas has positive
temperature contributions with negative to positive density contributions. 
The oxygen and carbon abundances increase in value as we move to the outside of the
manifold. The distribution on the left side, on the other hand, consists of the 
cold gas with low to medium densities. Here we can see that the ratio is mostly
impacted by the negative temperature contribution, combined with a positive to negative
density contribution. Similarly, the carbon and oxygen values are distributed in the radial
direction. Lastly, there is also a separated distribution on the bottom left; this again consists of
the high temperature, high density gas. We can again see that it is influenced by both the
carbon and oxygen content and that from the bottom left to the top right the ratio
increases as we go from low to high oxygen and carbon.

\subsubsection{\CNvHCN}
The manifold is displayed in \Cref{fig:results:CNvHCN:umap}. It is separated
into four distinct regions, the bottom left, left, right and top part. 
The bottom left region is strongly dominated by \HCN~, it contains mostly 
low temperature models, high carbon abundances and intermediate densities. 
The adjacent left distribution is dominated by higher density gas, but now shows
a clear split with respect to carbon abundance. Left of this split, the carbon abundance is higher, while on the right the carbon abundance is lower.
This split carries on into the right distribution of positive ratios with lower densities, with the top having low carbon abundance and bottom higher values. 
Lastly, there is a distribution of top right models with medium densities and
very low carbon abundances. These models have a positive ratio, which
is mostly explained by  the positive impact of low carbon, high oxygen and
medium temperature models.

\subsubsection{\HCOPvHCN}
The manifold for this ratio can be seen in \Cref{fig:results:HCOPvHCN:umap}. 
Across the horizontal axis, the density increases from left to right, whilst the ratios decrease. On the bottom left, there is a low density, low temperature
region, that has similar ratios as the top right distribution, but now explained
by a positive density impact and negative temperature impact. The top left, low
density region consists of \HCOP~ dominated gas, whilst the gas in
the right top is dominated by \HCN. The bottom right then contains a
region with high temperature gas and high densities, but neutral ratios as
the impacts sum to zero.

\subsubsection{\HNCvHCN}
We investigate the manifold for this ratio in \Cref{fig:results:HNCvHCN:umap}. It shows
no clear separation between the regions, but more of a continuous shift, with only
a \HCN~ dominated area in the bottom. This is in line
with the isomer chemistries being similar in nature. The large region spanning the rightmost part
of the plot is dominated by models with a small positive SHAP impact, resulting in a neutral ratio.
Then in the bottom we can see a region where the \HCN~ starts to dominate strongly. 
These models are low in carbon, with an associated negative carbon impact and a 
negative oxygen impact. This indicates that in order to get much \HNC~ depleted
models, we need low carbon abundances.
Lastly we move into the upper left part of the distribution, where the ratio
is again closer to being neutral. 

\subsubsection{\CSvSO}
The manifold, as can be seen in \Cref{fig:results:CSvSO:umap}, shows a complex but continuous
pattern, where the temperature impact largely varies across the vertical axis,
whilst the horizontal axis varies the carbon impact. There is a distinct
region in the center where the density impact is very positive, and
its values are low. In the top to middle distribution the ratios indicate a 
\SO~ deprived chemistry, with low carbon abundances. Moving to the bottom right part of the distribution,
we see the models with negative values, mostly driven by gas
of medium temperatures and high densities.

\subsubsection{\SIOvSO}
The UMAP manifold is displayed in \Cref{fig:results:SIOSO:umap}. 
This shows a distribution split between top and bottom regions with
mostly high and low temperatures, respectively. The left part of the manifold corresponds
to regions dominated by \SIO. In the rightmost high density region there is a distribution that
reaches negative values as the \SO~ is enhanced around 50$\;$K.

\subsubsection{\CSvCN}
The manifold, as can be seen in \Cref{fig:results:CSvCN:umap}, contains two
large distributions, with a smaller one in the bottom. The right distribution consists
of gas with a negative density contribution, creating values that are close
to zero, and even more negative on the edges, where the radiation field has a negative impact.
This shows that at low densities and high radiation fields, the impact of radiation field
becomes important. This can be attributed to the fact the \CN~ is related to photochemistry as shown 
in the \CNvHCN~ ratio before. We can also see that, indeed, the contribution
for the radiation field strongly depends on the density. As it increases at low densities,
it correctly lowers the ratio. The regions in the top are also dominated  by 
\CN~, but these models are related to a very negative carbon abundance impact instead.
With high densities and low carbon abundances, the ratio becomes negative. 

The distribution on the bottom, on the other hand, contains high carbon models, with
low temperatures and medium densities.
This results in a depletion of the \CN~chemistry. The region on the lower left contains 
a more broad distribution of models with generally positive density impact, high
densities and high temperatures, resulting in a similar but less extreme scenario.

 \newpage
\section{Discussion and conclusions} \label{sec:conclusion}
Overall, the results of this study point  to the physical parameters that are most important for chemistry, and that are therefore constrainable  with molecular observations.
Temperature and density are generally the most important features.
Therefore, observers that would like to use molecular observations to constrain other physical parameters need to first accurately probe temperature and density.
However, our study also suggests that not all the inspected physical parameters are relevant in the investigated molecular ratios.
For example, at these low temperatures and reduced metallicity, they cannot constrain the cosmic ray ionization rate, which hence would need to be estimated by other molecular species, or other methods.
This result was also found in Fontani et al. (2024), in which the technique presented here is used: the investigated molecular ratios are indeed not able to constrain the cosmic ray ionization rate in the interval of studied values ($\sim 10^{-14}-10^{-17}$ s$^{-1}$).
Some ratios are more sensitive than others to metallicity, and in particular to carbon abundance variations, as we will describe below.
Therefore, these ratios could also be used to test the metallicity gradients derived from observations.
For example, carbon and oxygen elemental  abundances are decreasing with the Galactocentric distance according to observations up to $\sim 14-16$~kpc (e.g. Arellano-Cordova+2020, Mendez-Delgado+2022).
These gradients are usually extrapolated to larger distances, where observations are lacking.
Our method allows to verify if such extrapolation is correct.
In fact, by measuring molecular ratios at Galactocentric distances larger than 16~kpc, as done for example in the CHEMOUT project, one can test if the measured molecular ratios are consistent with models in which elemental abundances are decreasing as predicted by the extrapolated gradients.

In this work we used a comprehensive grid of astrochemical models to understand the forward connection between the physical modeling parameters
and the observable line ratios. The nine ratios we choose to inspect
cover a large part of the physical parameter space, with the exception
of low temperature, high density regimes. We specifically focused on the
influence of lowering the initial elemental abundances of carbon
and oxygen. We used SHAP as a method to explain the
impact of each feature to the modeled ratio. This showed that the impact
of especially carbon and oxygen initial elemental abundances 
can be on par with temperature and density. 
We also showed that these SHAP explanations can be paired with UMAP to provide
insightful lower dimensional groupings for the models. 
We conclude the following about the molecular line ratios in this study:
\begin{enumerate}
\item The temperature and density are generally the most important features; Whereas the cosmic ray ionization is of low importance; this may be partially explained by the fact that the gas-grain model we used does not compute the temperature balance (and hence the temperature is independent from the cosmic rays); this isolates their effect only to the cosmic ray chemistry. 
\item Ratios such as \CNvHCN~and \HNCvHCN~can be sensitive to the initial carbon abundance, making them important for constraining the metallicity.
\item There is a distinct range of very negative \CNvHCN~ and \HNCvHCN~ models with a low carbon and high oxygen signature. 
\item The methanol based ratios depend largely on temperature and density, but in high temperature, high density regions, the carbon and even
oxygen abundances can have a large impact. 
\item Both the \HCO~ and \HCOP~based ratios depend little on the oxygen and carbon abundance in this parameter space.
\item The sulfur based ratios have a large dependence on
the carbon ratio, with distinct different regions of temperature and density,
allowing for effectively constraining the carbon ratio with observations.
\item None of the chosen ratios are particularly sensitive to the oxygen initial abundances, with \CSvCN~ having the largest oxygen dependence. 
\end{enumerate}
These analyses altogether can be used in order to inform new
forward model studies, but also the backward interpretation of
observations, by better informing the priors of Bayesian analysis.

\section*{Acknowledgements}
We thank the anonymous reviewer for their insightful comments and suggestions, which helped improve this manuscript.
G.V., F.F. and S.V. acknowledge support from the European Research Council (ERC) Advanced grant MOPPEX 833460. 
\bibliographystyle{aa} 
\bibliography{references.bib}

\begin{thebibliography}{67}
\expandafter\ifx\csname natexlab\endcsname\relax\def\natexlab#1{#1}\fi

\bibitem[{Akiba {et~al.}(2019)Akiba, Sano, Yanase, Ohta, \&
  Koyama}]{akibaOptunaNextgenerationHyperparameter2019}
Akiba, T., Sano, S., Yanase, T., Ohta, T., \& Koyama, M. 2019, Optuna: {{A
  Next-generation Hyperparameter Optimization Framework}}

\bibitem[{Bacmann \& Faure(2016)}]{bacmannOriginGasphaseHCO2016}
Bacmann, A. \& Faure, A. 2016, A\&A, 587, A130

\bibitem[{Bayet {et~al.}(2011)Bayet, Hartquist, Williams, Viti, Bell, \&
  Papadopoulos}]{bayetHowCosmicRays2011}
Bayet, E., Hartquist, T.~W., Williams, D.~A., {et~al.} 2011, Memorie della
  Societa Astronomica Italiana, 82, 893

\bibitem[{Bayet {et~al.}(2008)Bayet, Viti, Williams, \&
  Rawlings}]{bayetMolecularTracersHighMass2008}
Bayet, E., Viti, S., Williams, D.~A., \& Rawlings, J. M.~C. 2008, ApJ, 676, 978

\bibitem[{Bayet {et~al.}(2009)Bayet, Viti, Williams, Rawlings, \&
  Bell}]{bayetMolecularTracersPdrDominated2009a}
Bayet, E., Viti, S., Williams, D.~A., Rawlings, J. M.~C., \& Bell, T. 2009,
  ApJ, 696, 1466

\bibitem[{Behrens {et~al.}(2022)Behrens, Mangum, Holdship, Viti, Harada,
  Martin, Sakamoto, Muller, Tanaka, Nakanishi, {Herrero-Illana}, Yoshimura,
  Aladro, Colzi, Emig, Henkel, Huang, Humire, Meier, \&
  Rivilla}]{behrensTracingInterstellarHeating2022}
Behrens, E., Mangum, J.~G., Holdship, J., {et~al.} 2022, ApJ, 939, 119

\bibitem[{Bernal {et~al.}(2021)Bernal, Sephus, \&
  Ziurys}]{bernalMethanolEdgeGalaxy2021}
Bernal, J.~J., Sephus, C.~D., \& Ziurys, L.~M. 2021, ApJ, 922, 106

\bibitem[{Brown {et~al.}(1988)Brown, Charnley, \&
  Millar}]{brownModelChemistryHot1988}
Brown, P.~D., Charnley, S.~B., \& Millar, T.~J. 1988, MNRAS, 231, 409

\bibitem[{Butterworth {et~al.}(2022)Butterworth, Holdship, Viti, \&
  {Garc{\'i}a-Burillo}}]{butterworthUnderstandingIfMolecular2022}
Butterworth, J., Holdship, J., Viti, S., \& {Garc{\'i}a-Burillo}, S. 2022,
  A\&A, 667, A131

\bibitem[{Chen {et~al.}(2022)Chen, Hashimoto, Goto, Kim, Santos, On, Lu, \&
  Hsiao}]{chenUncloakingHiddenRepeating2022}
Chen, B.~H., Hashimoto, T., Goto, T., {et~al.} 2022, MNRAS, 509, 1227

\bibitem[{Chen {et~al.}(2019)Chen, Lundberg, \&
  Lee}]{chenExplainingModelsPropagating2019}
Chen, H., Lundberg, S., \& Lee, S.-I. 2019, Explaining {{Models}} by
  {{Propagating Shapley Values}} of {{Local Components}}

\bibitem[{Chen \& Guestrin(2016)}]{chenXGBoostScalableTree2016}
Chen, T. \& Guestrin, C. 2016, in Proceedings of the 22nd {{ACM SIGKDD
  International Conference}} on {{Knowledge Discovery}} and {{Data Mining}},
  785--794

\bibitem[{Codella \&
  Bachiller(1999)}]{codellaMolecularOutflowsIntermediatemass1999}
Codella, C. \& Bachiller, R. 1999, A\&A, 350, 659

\bibitem[{Colzi {et~al.}(2022)Colzi, Romano, Fontani, Rivilla, Bizzocchi,
  Beltran, Caselli, Elia, \& Magrini}]{colziCHEMOUTCHEMicalComplexity2022}
Colzi, L., Romano, D., Fontani, F., {et~al.} 2022, A\&A, 667, A151

\bibitem[{Esteban {et~al.}(2017)Esteban, Fang, {Garc{\'i}a-Rojas}, \& Toribio
  San~Cipriano}]{estebanRadialAbundanceGradient2017}
Esteban, C., Fang, X., {Garc{\'i}a-Rojas}, J., \& Toribio San~Cipriano, L.
  2017, MNRAS, 471, 987

\bibitem[{Fontani {et~al.}(2022{\natexlab{a}})Fontani, Colzi, Bizzocchi,
  Rivilla, Elia, Beltr{\'a}n, Caselli, Magrini, {S{\'a}nchez-Monge}, Testi, \&
  Romano}]{fontaniCHEMOUTCHEMicalComplexity2022}
Fontani, F., Colzi, L., Bizzocchi, L., {et~al.} 2022{\natexlab{a}}, A\&A, 660,
  A76

\bibitem[{Fontani {et~al.}(2022{\natexlab{b}})Fontani, Schmiedeke,
  {S{\'a}nchez-Monge}, Colzi, Elia, Rivilla, Beltr{\'a}n, Bizzocchi, Caselli,
  Magrini, \& Romano}]{fontaniCHEMOUTCHEMicalComplexity2022a}
Fontani, F., Schmiedeke, A., {S{\'a}nchez-Monge}, A., {et~al.}
  2022{\natexlab{b}}, A\&A, 664, A154

\bibitem[{Fontani {et~al.}(2024)Fontani, Vermari{\"e}n, Viti, Gigli, Colzi,
  Beltr{\'a}n, Caselli, Rivilla, \&
  {S{\'a}nchez-Monge}}]{fontaniCHEMOUTCHEMicalComplexity2024}
Fontani, F., Vermari{\"e}n, G., Viti, S., {et~al.} 2024, A\&A, 691, A180

\bibitem[{Gal {et~al.}(2021)Gal, {\"O}berg, Teague, Loomis, Law, Walsh, Bergin,
  Menard, Wilner, Andrews, Aikawa, Booth, Cataldi, Bergner, Bosman, Cleeves,
  Czekala, Furuya, Guzm{\'a}n, Huang, Ilee, Nomura, Qi, Schwarz, Tsukagoshi,
  Yamato, \& Zhang}]{galMoleculesALMAPlanetforming2021}
Gal, R.~L., {\"O}berg, K.~I., Teague, R., {et~al.} 2021, ApJ Supplement Series,
  257, 12

\bibitem[{{Garc{\'i}a-Burillo} {et~al.}(2010){Garc{\'i}a-Burillo}, Usero,
  Fuente, {Mart{\'i}n-Pintado}, Boone, Aalto, Krips, Neri, Schinnerer, \&
  Tacconi}]{garcia-burilloMolecularGasChemistry2010}
{Garc{\'i}a-Burillo}, S., Usero, A., Fuente, A., {et~al.} 2010, A\&A, 519, A2

\bibitem[{Grassi {et~al.}(2025)Grassi, Padovani, Galli, Vaytet, Jensen,
  Redaelli, Spezzano, Bovino, \&
  Caselli}]{grassiMappingSyntheticObservations2025}
Grassi, T., Padovani, M., Galli, D., {et~al.} 2025, Mapping {{Synthetic
  Observations}} to {{Prestellar Core Models}}: {{An Interpretable Machine
  Learning Approach}}

\bibitem[{Hacar {et~al.}(2020)Hacar, Bosman, \&
  Van~Dishoeck}]{hacarHCNtoHNCIntensityRatio2020}
Hacar, A., Bosman, A.~D., \& Van~Dishoeck, E.~F. 2020, A\&A, 635, A4

\bibitem[{Harada {et~al.}(2024{\natexlab{a}})Harada, Meier, Mart{\'i}n, Muller,
  Sakamoto, Saito, Gorski, Henkel, Tanaka, Mangum, Aalto, Aladro, Bouvier,
  Colzi, Emig, {Herrero-Illana}, Huang, Kohno, K{\"o}nig, Nakanishi, Nishimura,
  Takano, Rivilla, Viti, Watanabe, {van der Werf}, \&
  Yoshimura}]{haradaALCHEMIAtlasPrincipal2024}
Harada, N., Meier, D.~S., Mart{\'i}n, S., {et~al.} 2024{\natexlab{a}}, The
  {{ALCHEMI}} Atlas: Principal Component Analysis Reveals Starburst Evolution
  in {{NGC}} 253

\bibitem[{Harada {et~al.}(2024{\natexlab{b}})Harada, Saito, Nishimura,
  Watanabe, \& Sakamoto}]{haradaTemperatureFUVTracer2024a}
Harada, N., Saito, T., Nishimura, Y., Watanabe, Y., \& Sakamoto, K.
  2024{\natexlab{b}}, A Temperature or {{FUV}} Tracer? {{The HNC}}/{{HCN}}
  Ratio in {{M83}} on the {{GMC}} Scale

\bibitem[{Hastie {et~al.}(2009)Hastie, Tibshirani, \&
  Friedman}]{hastieElementsStatisticalLearning2009}
Hastie, T., Tibshirani, R., \& Friedman, J. 2009, The {{Elements}} of
  {{Statistical Learning}}, Springer {{Series}} in {{Statistics}} (New York,
  NY: Springer)

\bibitem[{Herbst \& van Dishoeck(2009)}]{herbstComplexOrganicInterstellar2009}
Herbst, E. \& van Dishoeck, E.~F. 2009, Annual Review of A\&A, 47, 427

\bibitem[{Heyl {et~al.}(2023{\natexlab{a}})Heyl, Butterworth, \&
  Viti}]{heylUnderstandingMolecularAbundances2023}
Heyl, J., Butterworth, J., \& Viti, S. 2023{\natexlab{a}}, MNRAS, 526, 404

\bibitem[{Heyl {et~al.}(2023{\natexlab{b}})Heyl, Viti, \&
  Vermari{\"e}n}]{heylStatisticalMachineLearning2023}
Heyl, J., Viti, S., \& Vermari{\"e}n, G. 2023{\natexlab{b}}, Faraday
  Discussions, 245, 569

\bibitem[{Holdship {et~al.}(2017)Holdship, Viti, {Jim{\'e}nez-Serra},
  Makrymallis, \& Priestley}]{holdshipUCLCHEMGasgrainChemical2017}
Holdship, J., Viti, S., {Jim{\'e}nez-Serra}, I., Makrymallis, A., \& Priestley,
  F. 2017, The Astronomical Journal, 154, 38

\bibitem[{Indriolo {et~al.}(2007)Indriolo, Geballe, Oka, \&
  McCall}]{indrioloH3DiffuseInterstellar2007}
Indriolo, N., Geballe, T.~R., Oka, T., \& McCall, B.~J. 2007, ApJ, 671, 1736

\bibitem[{James {et~al.}(2021)James, Viti, {Yusef-Zadeh}, Royster, \&
  Wardle}]{jamesRevealingPhysicalConditions2021a}
James, T.~A., Viti, S., {Yusef-Zadeh}, F., Royster, M., \& Wardle, M. 2021,
  ApJ, 916, 69

\bibitem[{Kane {et~al.}(2023)Kane, Hawkins, \&
  Maas}]{kaneHuntGalacticFossils2023}
Kane, S., Hawkins, K., \& Maas, Z. 2023, 241, 208.11

\bibitem[{K{\"o}nig {et~al.}(2018)K{\"o}nig, Aalto, Muller, Gallagher, Beswick,
  Varenius, J{\"u}tte, Krips, \& Adamo}]{konigMajorImpactMinor2018}
K{\"o}nig, S., Aalto, S., Muller, S., {et~al.} 2018, A\&A, 615, A122

\bibitem[{Lamb {et~al.}(2019)Lamb, Malhotra, Vlontzos, Wagstaff,
  G{\"u}nes~Baydin, Bhiwandiwalla, Gal, Kalaitzis, Reina, \&
  Bhatt}]{lambCorrelationAuroralDynamics2019}
Lamb, K., Malhotra, G., Vlontzos, A., {et~al.} 2019, Correlation of {{Auroral
  Dynamics}} and {{GNSS Scintillation}} with an {{Autoencoder}}

\bibitem[{Li {et~al.}(2015)Li, Wang, Zhu, Zhang, \&
  Li}]{liSulfurbearingMoleculesMassive2015}
Li, J., Wang, J., Zhu, Q., Zhang, J., \& Li, D. 2015, ApJ, 802, 40

\bibitem[{Lundberg \& Lee(2017)}]{lundbergUnifiedApproachInterpreting2017}
Lundberg, S. \& Lee, S.-I. 2017, A {{Unified Approach}} to {{Interpreting Model
  Predictions}}

\bibitem[{Lundberg {et~al.}(2020)Lundberg, Erion, Chen, DeGrave, Prutkin, Nair,
  Katz, Himmelfarb, Bansal, \& Lee}]{lundbergLocalExplanationsGlobal2020}
Lundberg, S.~M., Erion, G., Chen, H., {et~al.} 2020, Nature Machine
  Intelligence, 2, 56

\bibitem[{McInnes {et~al.}(2020)McInnes, Healy, \&
  Melville}]{mcinnesUMAPUniformManifold2020}
McInnes, L., Healy, J., \& Melville, J. 2020, {{UMAP}}: {{Uniform Manifold
  Approximation}} and {{Projection}} for {{Dimension Reduction}}

\bibitem[{{M{\'e}ndez-Delgado} {et~al.}(2022){M{\'e}ndez-Delgado}, Amayo,
  {Arellano-C{\'o}rdova}, Esteban, {Garc{\'i}a-Rojas}, Carigi, \&
  {Delgado-Inglada}}]{mendez-delgadoGradientsChemicalAbundances2022}
{M{\'e}ndez-Delgado}, J.~E., Amayo, A., {Arellano-C{\'o}rdova}, K.~Z., {et~al.}
  2022, MNRAS, 510, 4436

\bibitem[{Milam {et~al.}(2005)Milam, Savage, Brewster, Ziurys, \&
  Wyckoff}]{milam1213Isotope2005}
Milam, S.~N., Savage, C., Brewster, M.~A., Ziurys, L.~M., \& Wyckoff, S. 2005,
  ApJ, 634, 1126

\bibitem[{Millar {et~al.}(1991)Millar, Bennett, Rawlings, Brown, \&
  Charnley}]{millarGasPhaseReactions1991}
Millar, T.~J., Bennett, A., Rawlings, J. M.~C., Brown, P.~D., \& Charnley,
  S.~B. 1991, A\&A Supplement Series, 87, 585

\bibitem[{Molnar(2022)}]{molnarInterpretableMachineLearning2022}
Molnar, C. 2022, Interpretable Machine Learning: A Guide for Making Black Box
  Models Explainable, second edition edn. (Munich, Germany: Christoph Molnar)

\bibitem[{Pearson(1901)}]{pearsonLIIILinesPlanes1901}
Pearson, K. 1901, The London, Edinburgh, and Dublin Philosophical Magazine and
  Journal of Science, 2, 559

\bibitem[{Pe{\~n}aloza {et~al.}(2018)Pe{\~n}aloza, Clark, Glover, \&
  Klessen}]{penalozaCOLineRatios2018}
Pe{\~n}aloza, C.~H., Clark, P.~C., Glover, S. C.~O., \& Klessen, R.~S. 2018,
  MNRAS, 475, 1508

\bibitem[{Ramos {et~al.}(2024)Ramos, Plaza, {Navarro-Almaida},
  {Rivi{\`e}re-Marichalar}, Wakelam, \& Fuente}]{ramosFastNeuralEmulator2024a}
Ramos, A.~A., Plaza, C.~W., {Navarro-Almaida}, D., {et~al.} 2024, MNRAS, 531,
  4930

\bibitem[{Ribeiro {et~al.}(2016)Ribeiro, Singh, \&
  Guestrin}]{ribeiroWhyShouldTrust2016}
Ribeiro, M.~T., Singh, S., \& Guestrin, C. 2016, "{{Why Should I Trust You}}?":
  {{Explaining}} the {{Predictions}} of {{Any Classifier}}

\bibitem[{Rollig {et~al.}(2007)Rollig, Abel, Bell, Bensch, Black, Ferland,
  Jonkheid, Kamp, Kaufman, Bourlot, Petit, Meijerink, Chirivella, Ossenkopf,
  Roueff, Shaw, Sternberg, \& Stutzki}]{rolligPDRCodeComparisonStudy2007}
Rollig, M., Abel, N.~P., Bell, T., {et~al.} 2007

\bibitem[{Ruaud {et~al.}(2016)Ruaud, Wakelam, \&
  Hersant}]{ruaudGasGrainChemical2016a}
Ruaud, M., Wakelam, V., \& Hersant, F. 2016, MNRAS, 459, 3756

\bibitem[{Sabatini {et~al.}(2021)Sabatini, Bovino, Giannetti, Grassi, Brand,
  Schisano, Wyrowski, Leurini, \&
  Menten}]{sabatiniEstablishingEvolutionaryTimescales2021}
Sabatini, G., Bovino, S., Giannetti, A., {et~al.} 2021, A\&A, 652, A71

\bibitem[{Scholbeck {et~al.}(2020)Scholbeck, Molnar, Heumann, Bischl, \&
  Casalicchio}]{scholbeckSamplingInterventionPrediction2020}
Scholbeck, C.~A., Molnar, C., Heumann, C., Bischl, B., \& Casalicchio, G. 2020,
  1167, 205

\bibitem[{Semenov {et~al.}(2018)Semenov, Favre, Fedele, Guilloteau, Teague,
  Henning, Dutrey, Chapillon, Hersant, \&
  Pi{\'e}tu}]{semenovChemistryDisksXI2018}
Semenov, D., Favre, C., Fedele, D., {et~al.} 2018, A\&A, 617, A28

\bibitem[{Sewi{\l}o {et~al.}(2018)Sewi{\l}o, Indebetouw, Charnley, Zahorecz,
  Oliveira, {van Loon}, Ward, Chen, Wiseman, Fukui, Kawamura, Meixner, Onishi,
  \& Schilke}]{sewiloDetectionHotCores2018}
Sewi{\l}o, M., Indebetouw, R., Charnley, S.~B., {et~al.} 2018, ApJ Letters,
  853, L19

\bibitem[{Sewi{\l}o {et~al.}(2022)Sewi{\l}o, Karska, Kristensen, Charnley,
  Chen, Oliveira, Cordiner, Wiseman, {S{\'a}nchez-Monge}, {van Loon},
  Indebetouw, Schilke, \&
  {Garcia-Berrios}}]{sewiloDetectionDeuteratedWater2022}
Sewi{\l}o, M., Karska, A., Kristensen, L.~E., {et~al.} 2022, ApJ, 933, 64

\bibitem[{Shapley \& Shubik(1971)}]{shapleyAssignmentGameCore1971}
Shapley, L.~S. \& Shubik, M. 1971, International Journal of Game Theory, 1, 111

\bibitem[{Shimonishi {et~al.}(2021)Shimonishi, Izumi, Furuya, \&
  Yasui}]{shimonishiDetectionHotMolecular2021}
Shimonishi, T., Izumi, N., Furuya, K., \& Yasui, C. 2021, ApJ, 922, 206

\bibitem[{Shimonishi {et~al.}(2023)Shimonishi, Tanaka, Zhang, \&
  Furuya}]{shimonishiDetectionHotMolecular2023}
Shimonishi, T., Tanaka, K. E.~I., Zhang, Y., \& Furuya, K. 2023, ApJ Letters,
  946, L41

\bibitem[{Sobol'(1967)}]{sobolDistributionPointsCube1967}
Sobol', I.~M. 1967, USSR Computational Mathematics and Mathematical Physics, 7,
  86

\bibitem[{Spezzano {et~al.}(2020)Spezzano, Caselli, Pineda, Bizzocchi,
  Prudenzano, \& Nagy}]{spezzanoDistributionMethanolCyclopropenylidene2020}
Spezzano, S., Caselli, P., Pineda, J.~E., {et~al.} 2020, A\&A, 643, A60

\bibitem[{Tafalla {et~al.}(2021)Tafalla, Usero, \&
  Hacar}]{tafallaCharacterizingLineEmission2021}
Tafalla, M., Usero, A., \& Hacar, A. 2021, A\&A, 646, A97

\bibitem[{Usero {et~al.}(2004)Usero, {Garc{\'i}a-Burillo}, Fuente,
  {Mart{\'i}n-Pintado}, \&
  {Rodr{\'i}guez-Fern{\'a}ndez}}]{useroMolecularGasChemistry2004}
Usero, A., {Garc{\'i}a-Burillo}, S., Fuente, A., {Mart{\'i}n-Pintado}, J., \&
  {Rodr{\'i}guez-Fern{\'a}ndez}, N.~J. 2004, A\&A, 419, 897

\bibitem[{{van der Maaten} \&
  Hinton(2008)}]{vandermaatenVisualizingHighDimensionalData2008}
{van der Maaten}, L. \& Hinton, G. 2008, Journal of Machine Learning Research,
  9, 2579

\bibitem[{Viti \& Williams(1999)}]{vitiTimedependentEvaporationIcy1999a}
Viti, S. \& Williams, D.~A. 1999, MNRAS, 305, 755

\bibitem[{Wakelam {et~al.}(2010)Wakelam, Herbst, Le~Bourlot, Hersant, Selsis,
  \& Guilloteau}]{wakelamSensitivityAnalysesDense2010}
Wakelam, V., Herbst, E., Le~Bourlot, J., {et~al.} 2010, A\&A, 517, A21

\bibitem[{Wang {et~al.}(2022)Wang, Qi, Li, \& Wu}]{wangCN21CS2022}
Wang, J., Qi, C., Li, S., \& Wu, J. 2022, ApJ, 937, 120

\bibitem[{Williams(1998)}]{williamsIntroductoryLectureFrontiers1998}
Williams, D.~A. 1998, Faraday Discussions, 109, 1

\bibitem[{Wilson {et~al.}(2023)Wilson, Bemis, Ledger, \&
  Klimi}]{wilsonNearlyConstantCN2023}
Wilson, C.~D., Bemis, A., Ledger, B., \& Klimi, O. 2023, MNRAS, 521, 717

\bibitem[{Woods {et~al.}(2012)Woods, Kelly, Viti, Slater, Brown, Puletti,
  Burke, \& Raza}]{woodsFormationGlycolaldehydeDense2012a}
Woods, P.~M., Kelly, G., Viti, S., {et~al.} 2012, ApJ, 750, 19

\end{thebibliography}

\newpage
\onecolumn
\begin{appendix}
\section{Elemental abundances}
The initial elemental abundances used in UCLCHEM and the depletion of the initial carbon and oyxgen abundances are listed in \Cref{tab:elemental_abundances}.
\label{app:elemental_abundances}
\begin{table}[ht!]
    \caption{Initial elemental abundances used in UCLCHEM. The number density of hydrogen nuclei is $n_{\mathrm{H},nuclei} = n_\mathrm{H} + 2 n_{\mathrm{H}_2}$.}

    \centering
    \begin{tabular}{|c|c|c|}
        \hline
        Element & Symbol & $x_i=n_i/n_{\mathrm{H},nuclei}$\\
        \hline
        Atomic hydrogen & H & $0.5$ \\
        Molecular hydrogen & H$_2$ & $0.5$ \\
        Helium & He & $0.1$ \\
        Carbon & C & $(0.05\text{ to } 1) \times 1.77 \times 10^{-4}$ \\
        Oxygen & O & $(0.05\text{ to } 1) \times 3.34 \times 10^{-4}$ \\
        Nitrogen & N & $6.18 \times 10^{-5}$ \\
        Sulfur & S & $3.51 \times 10^{-6}$ \\
        Magnesium & Mg & $2.256 \times 10^{-6}$ \\
        Silicon & Si & $1.78 \times 10^{-6}$ \\
        Chlorine & Cl & $3.39 \times 10^{-8}$ \\
        Phosphorus & P & $7.78 \times 10^{-8}$ \\
        Iron & Fe & $2.01 \times 10^{-7}$ \\
        Fluorine & F & $3.6 \times 10^{-8}$ \\
        \hline
    \end{tabular}
    \label{tab:elemental_abundances}
\end{table}

\FloatBarrier
\section{Hyperparameter optimization}
\label{app:optuna}
The best hyperparameters for the xgboost regression forests \citep{chenXGBoostScalableTree2016} after 500 trials using Optuna \citep{akibaOptunaNextgenerationHyperparameter2019} can be found in \Cref{app:hparams}
\begin{table*}[h!]
\caption{The ranges for each hyperparameter and the best configuration found after 500 trials with Optuna \citep{akibaOptunaNextgenerationHyperparameter2019}.}
\begin{tabular}{lllllll}
\toprule
 {}& \texttt{lambda} & \texttt{alpha} & \texttt{subsample} & \texttt{colsample}\_\texttt{bytree} & \texttt{max\_depth} & \texttt{n}\_\texttt{estimators} \\
$\mathbf{minimum}$ & $10^{-8}$ & $10^{-8}$ & 0.2 & 0.2 & 3 & 1 \\
$\mathbf{maximum}$ & 1.0 & 1.0 & 1.0  & 1.0 & 15 & 1000 \\ 
\midrule
\HIICOvCHIIIOH & $6.7\times10^{-1}$ & $4.2\times10^{-1}$ & $9.2\times10^{-1}$ & $8.6\times10^{-1}$ & $1.2\times10^{1}$ & $8.8\times10^{2}$ \\
\HCOvCHIIIOH & $7.2\times10^{-1}$ & $9.9\times10^{-6}$ & $9.7\times10^{-1}$ & $9.4\times10^{-1}$ & $7.0\times10^{0}$ & $5.5\times10^{2}$ \\
\CIIIHIIvCHIIIOH & $6.1\times10^{-1}$ & $1.0\times10^{-7}$ & $9.1\times10^{-1}$ & $9.5\times10^{-1}$ & $9.0\times10^{0}$ & $2.9\times10^{2}$ \\
\CNvHCN & $1.9\times10^{-5}$ & $8.9\times10^{-4}$ & $8.8\times10^{-1}$ & $8.7\times10^{-1}$ & $1.1\times10^{1}$ & $7.4\times10^{2}$ \\
\HCOPvHCN & $5.9\times10^{-6}$ & $1.8\times10^{-4}$ & $9.4\times10^{-1}$ & $9.7\times10^{-1}$ & $8.0\times10^{0}$ & $5.9\times10^{2}$ \\
\HNCvHCN & $1.7\times10^{-4}$ & $8.5\times10^{-1}$ & $8.2\times10^{-1}$ & $8.8\times10^{-1}$ & $1.3\times10^{1}$ & $5.8\times10^{2}$ \\
\CSvSO & $1.8\times10^{-4}$ & $8.4\times10^{-1}$ & $8.0\times10^{-1}$ & $8.4\times10^{-1}$ & $7.0\times10^{0}$ & $9.8\times10^{2}$ \\
\SIOvSO & $6.2\times10^{-8}$ & $2.6\times10^{-1}$ & $9.2\times10^{-1}$ & $8.6\times10^{-1}$ & $1.0\times10^{1}$ & $8.1\times10^{2}$ \\
\CSvCN & $9.7\times10^{-1}$ & $6.1\times10^{-1}$ & $9.9\times10^{-1}$ & $9.4\times10^{-1}$ & $1.0\times10^{1}$ & $1.4\times10^{2}$ \\
\bottomrule
\end{tabular}
\label{app:hparams}
\end{table*}
\FloatBarrier
\section{UMAP plots}
This appendix contains all the remaining UMAP plots for \CIIIHII, \HCOvCHIIIOH, \CNvHCN, \HCOPvHCN, \HNCvHCN, \CSvSO, \SIOvSO, \CSvCN.
\graphicspath{{figures/time_1e5/umap}}
\begin{figure*}
    \centering
    \includegraphics[width=\linewidth]{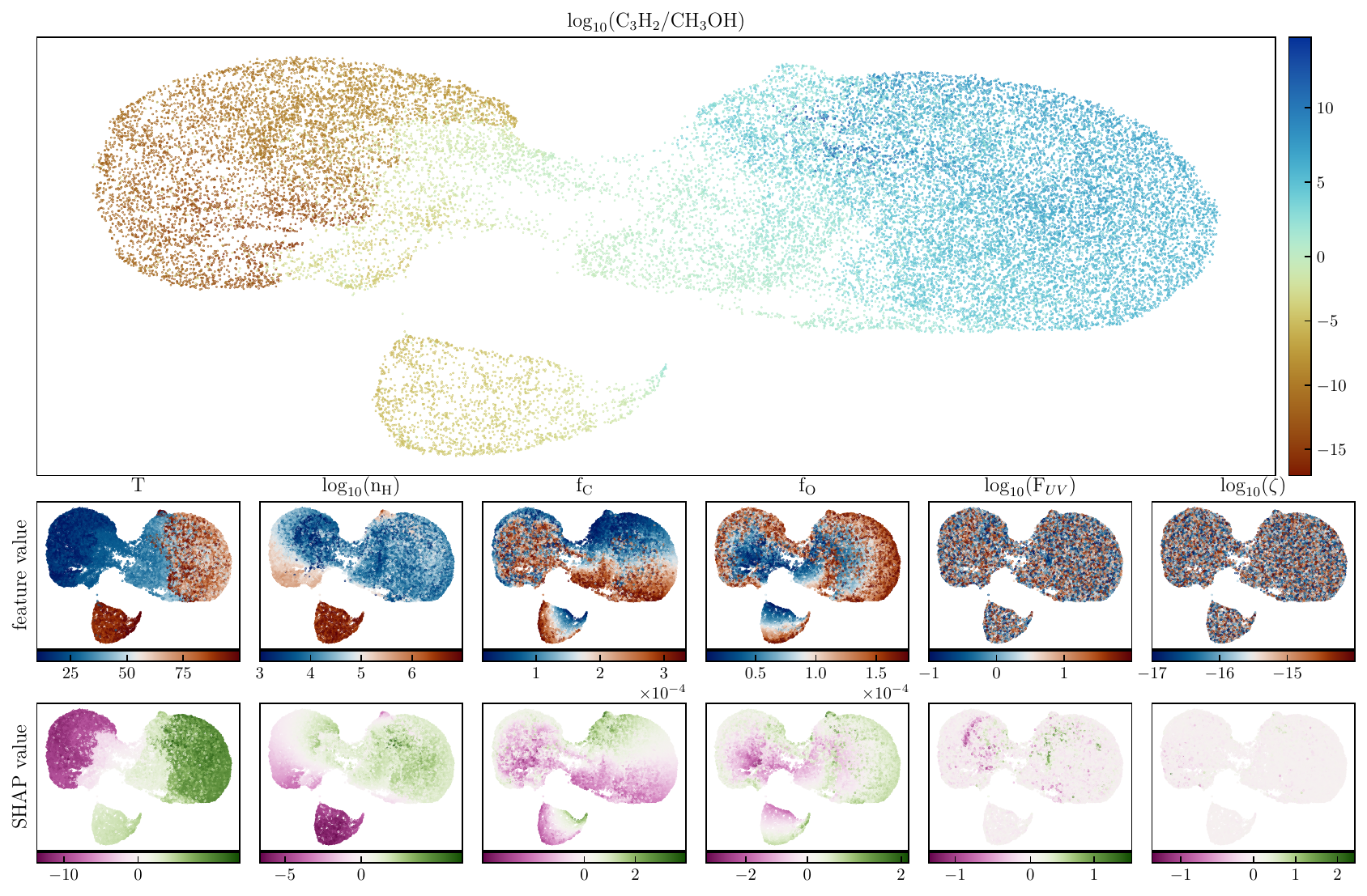}
    \caption{The \CIIIHIIvCHIIIOH~ratio, plotted on the manifold. The manifold is separated into three broad
    regions.}
    \label{fig:results:c3h2ch3oh:umap}
\end{figure*}
\begin{figure*}
    \centering
    \includegraphics[width=\linewidth]{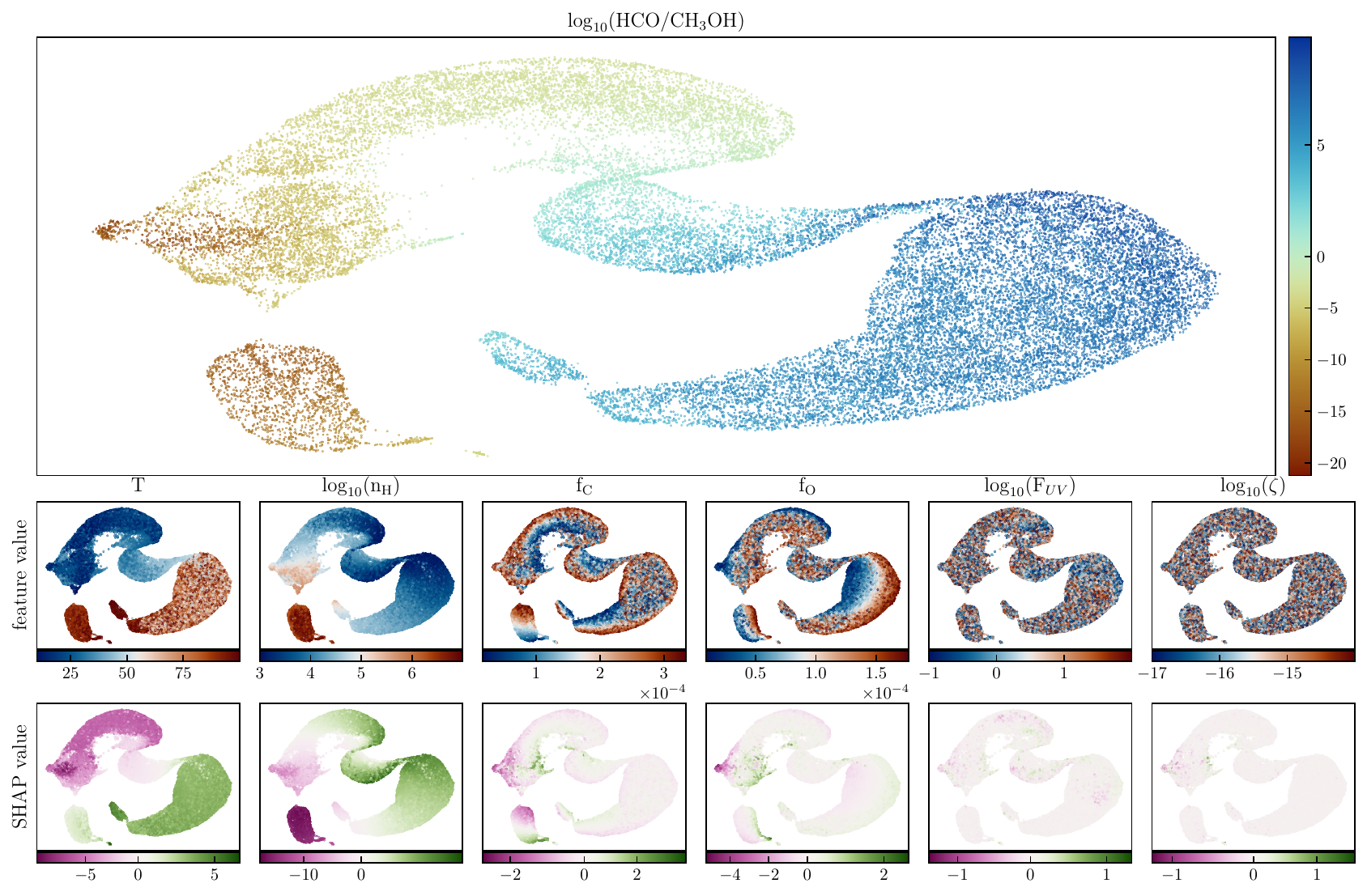} 
    \caption{The \HCOvCHIIIOH~ratio, plotted on the manifold. The manifold is a smooth continuous distribution with one separated lobe.} 
    \label{fig:results:hcoh3oh:umap}
\end{figure*}
\begin{figure*}
    \centering
    \includegraphics[width=\linewidth]{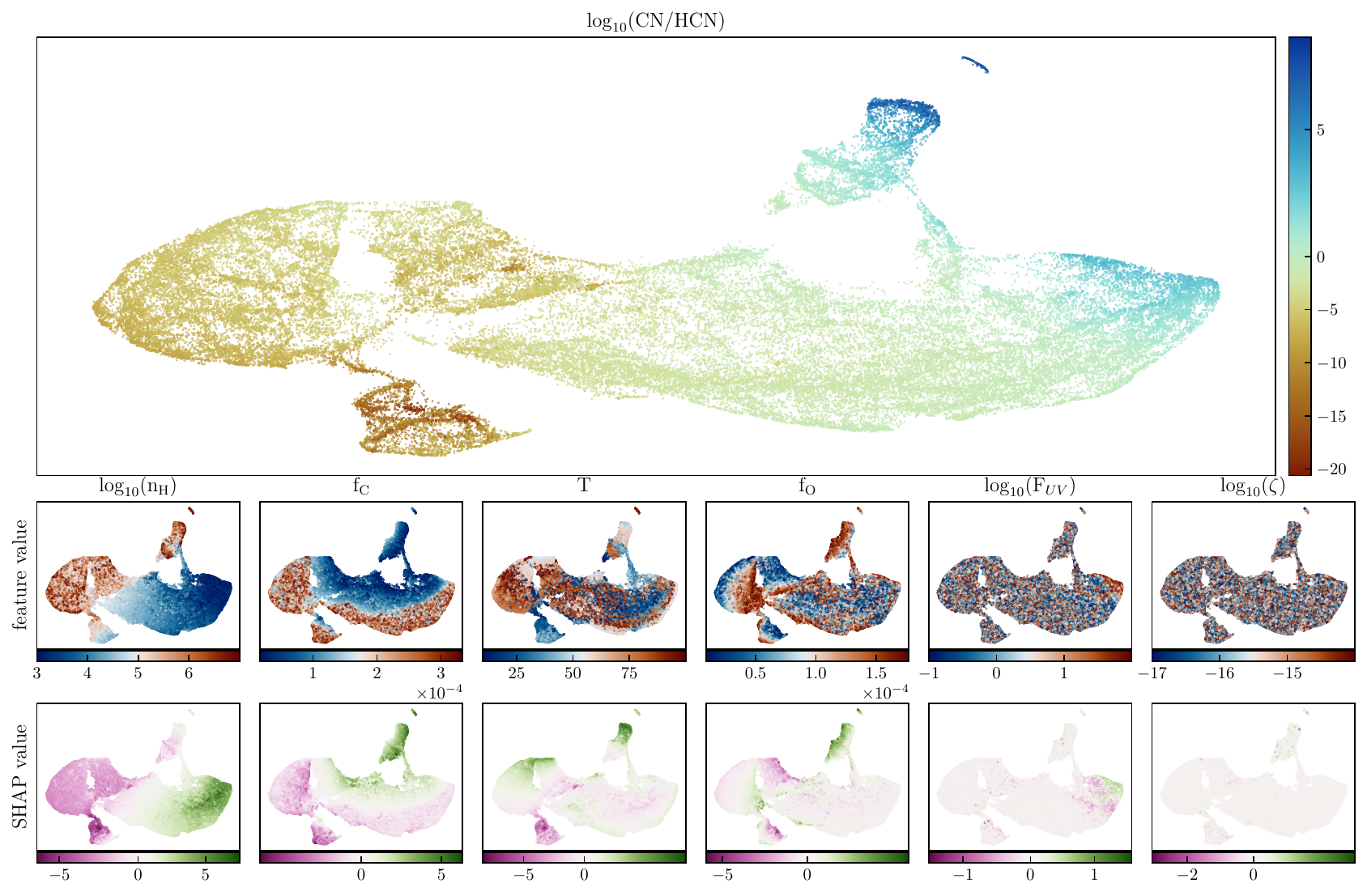}
    \caption{The \CNvHCN~ratio, plotted on the manifold. It shows a smooth manifold with a gradient in ratio from top to bottom.}    
    \label{fig:results:CNvHCN:umap}
\end{figure*}

\begin{figure*}
    \centering
    \includegraphics[width=\linewidth]{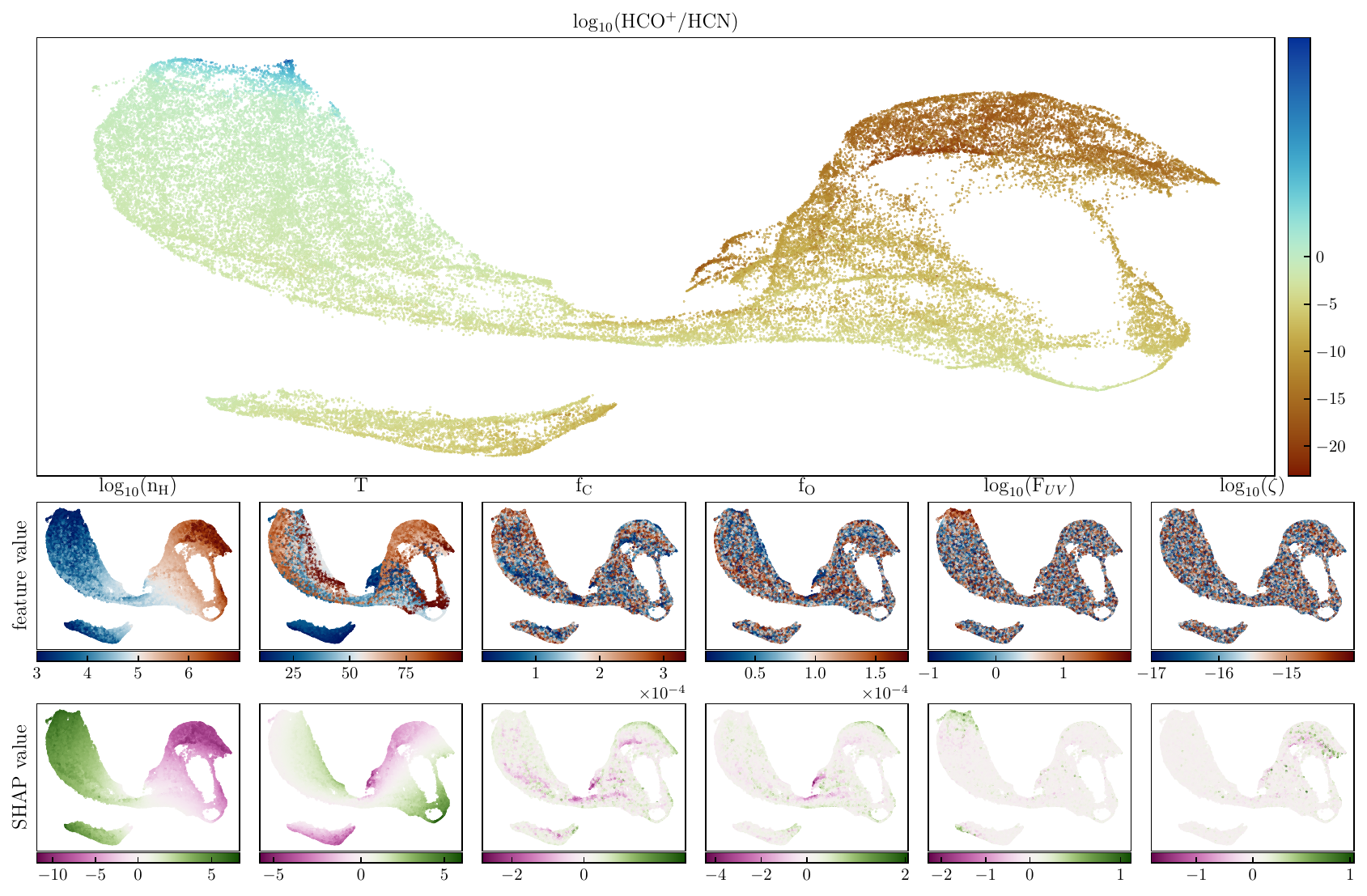}
    \caption{The \HCOPvHCN~ratio plotted over a relatively smooth manifold. 
    It shows an elongated manifold with a separate distribution 
    in the lower left corner.}    
    \label{fig:results:HCOPvHCN:umap}
\end{figure*}\begin{figure*}
    \centering
    \includegraphics[width=\linewidth]{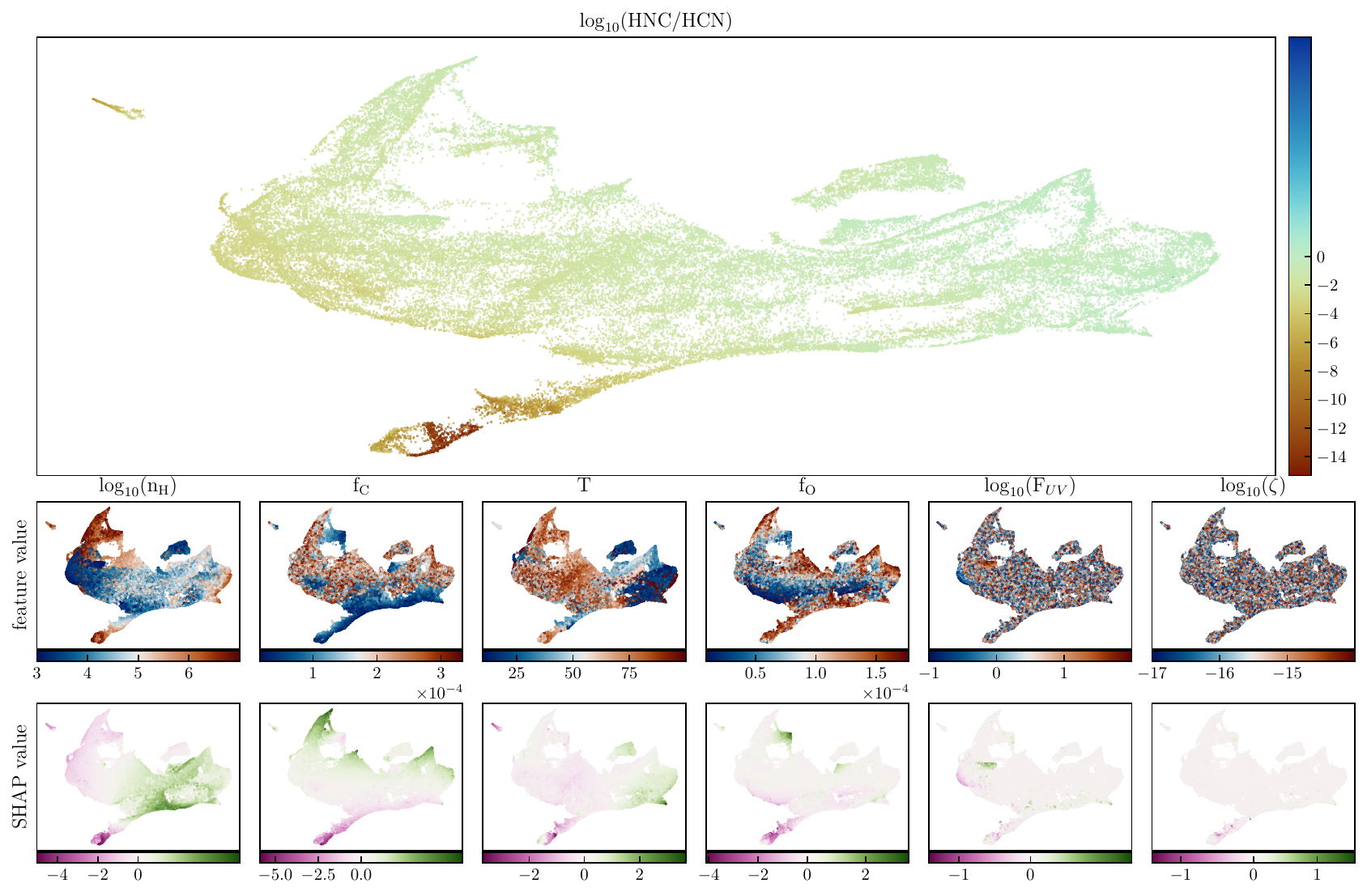}
    \caption{The \HNCvHCN~ratio plotted on the manifold. The manifold
    separates a general region with equal ratio and a lower lobe
    with \HCN~enhancement.}
    \label{fig:results:HNCvHCN:umap}
\end{figure*}\begin{figure*}
    \centering
    \includegraphics[width=\linewidth]{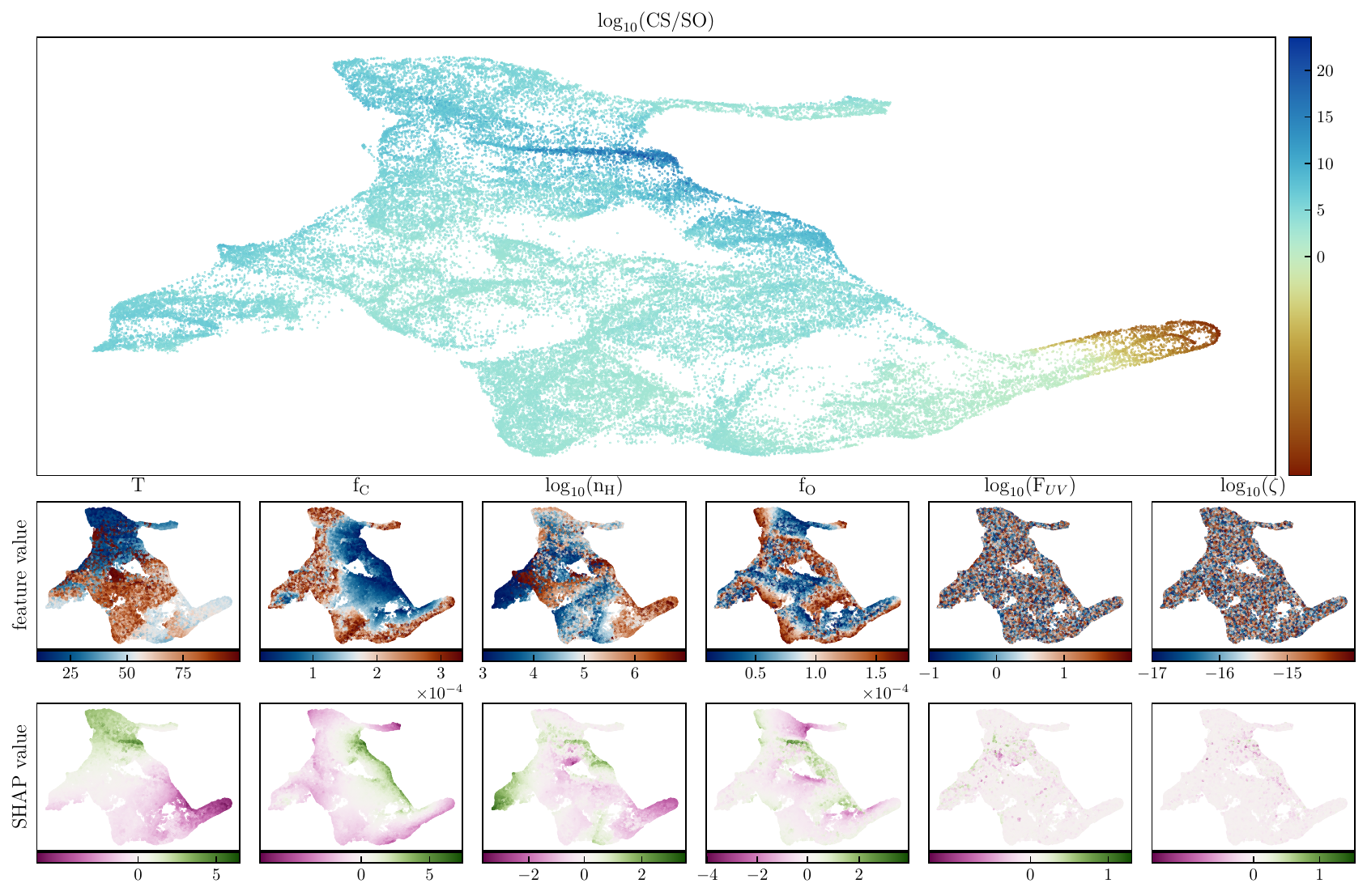}
    \caption{The \CSvSO~ratio again shows no clear separation of regions
    on the manifold. With a tail in the lower right containing 
    \SO enhanced models.} 
    \label{fig:results:CSvSO:umap} 
\end{figure*}\begin{figure*}
    \centering
    \includegraphics[width=\linewidth]{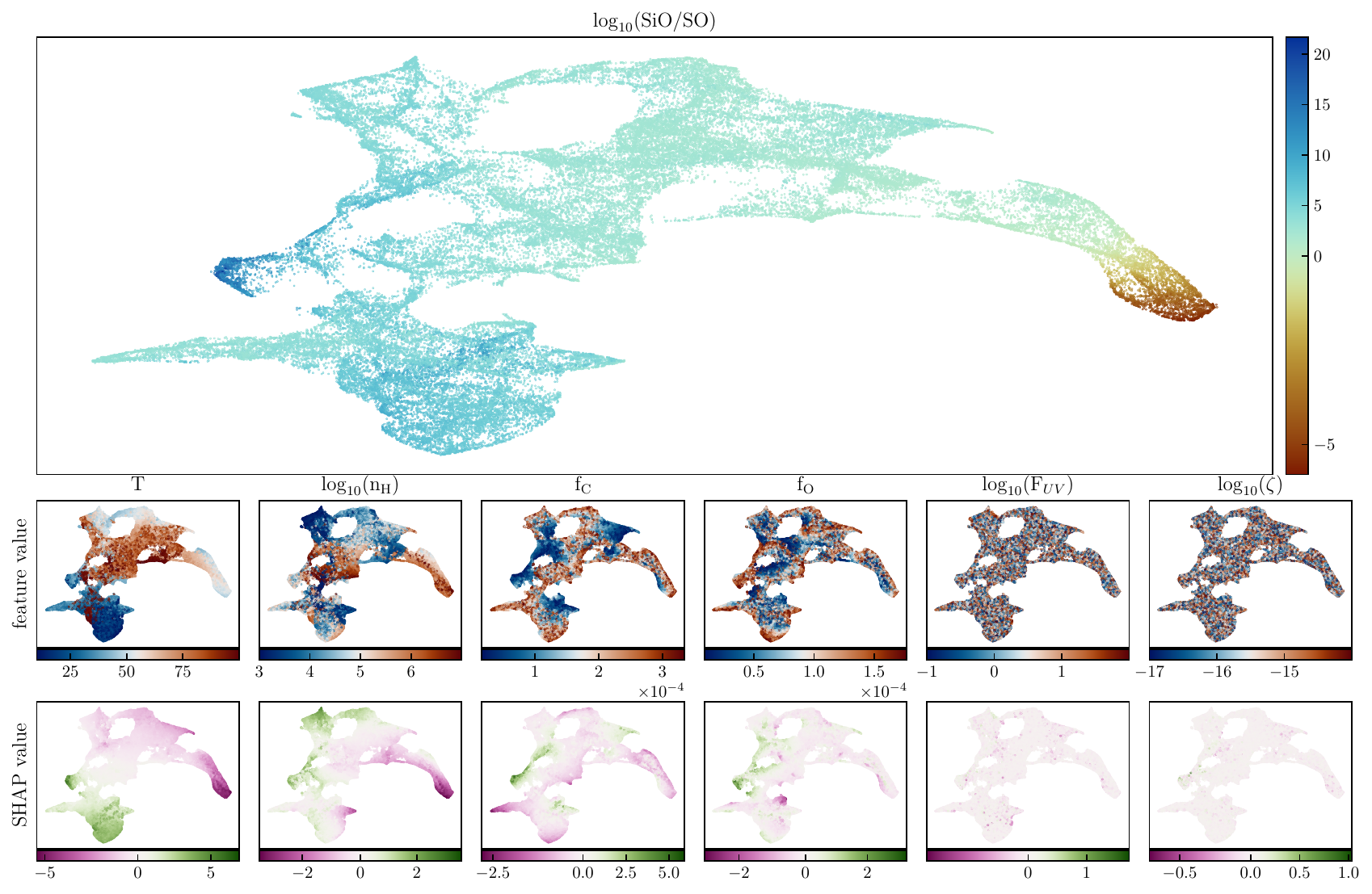}
    \caption{The \SIOvSO~ratio plotted on the manifold. It separates into two
    broad regions, influenced by temperature.}
    \label{fig:results:SIOSO:umap}
\end{figure*}
\graphicspath{{figures/time_1e5/umap}}
\begin{figure*}
    \centering
    \includegraphics[width=\linewidth]{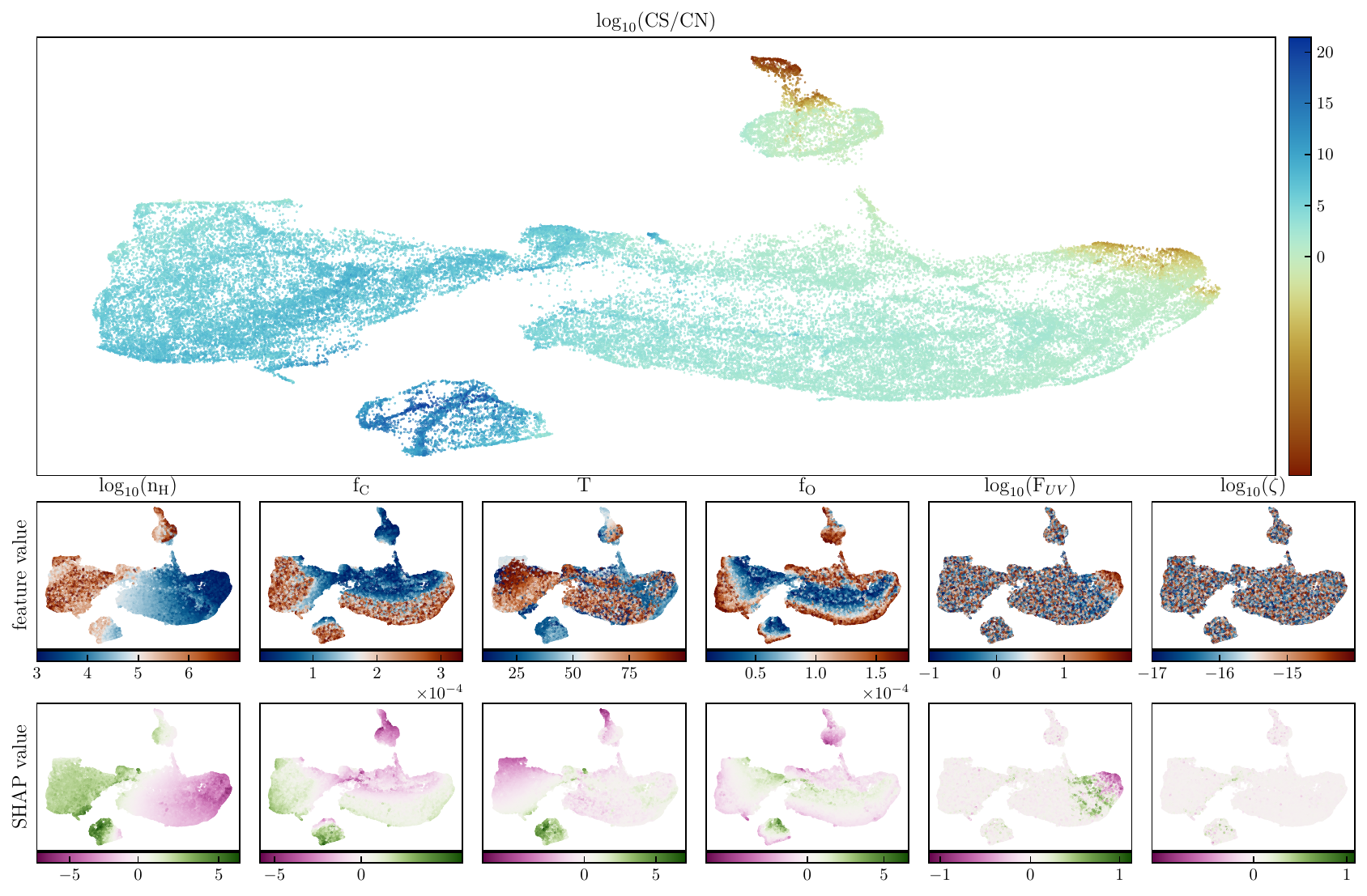}
    \caption{The \CSvCN~ratio plotted on the manifold shows a separation between four global regions.}    
    \label{fig:results:CSvCN:umap}
\end{figure*}
\FloatBarrier

\section{Enumerator and denominator of ratios in density-temperature space}
\label{app:enum_denom}
\begin{figure*}[h!]
  \includegraphics[width=\linewidth]{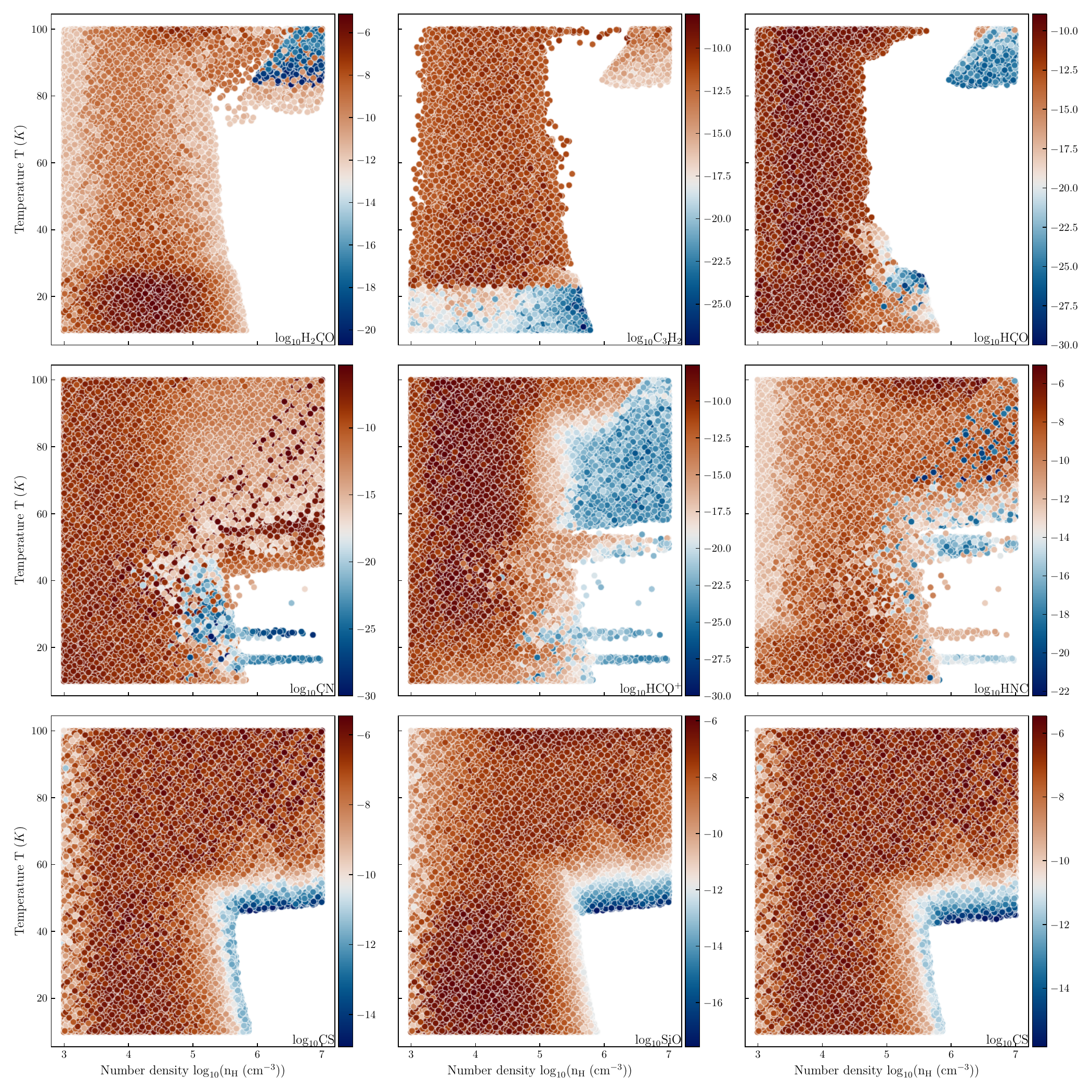}
  \caption{The enumerator for the ratios as a function of density and temperature including
  the observational limit.}  
  \label{fig:app:enum_nh_t}
\end{figure*}

\begin{figure*}
  \includegraphics[width=\linewidth]{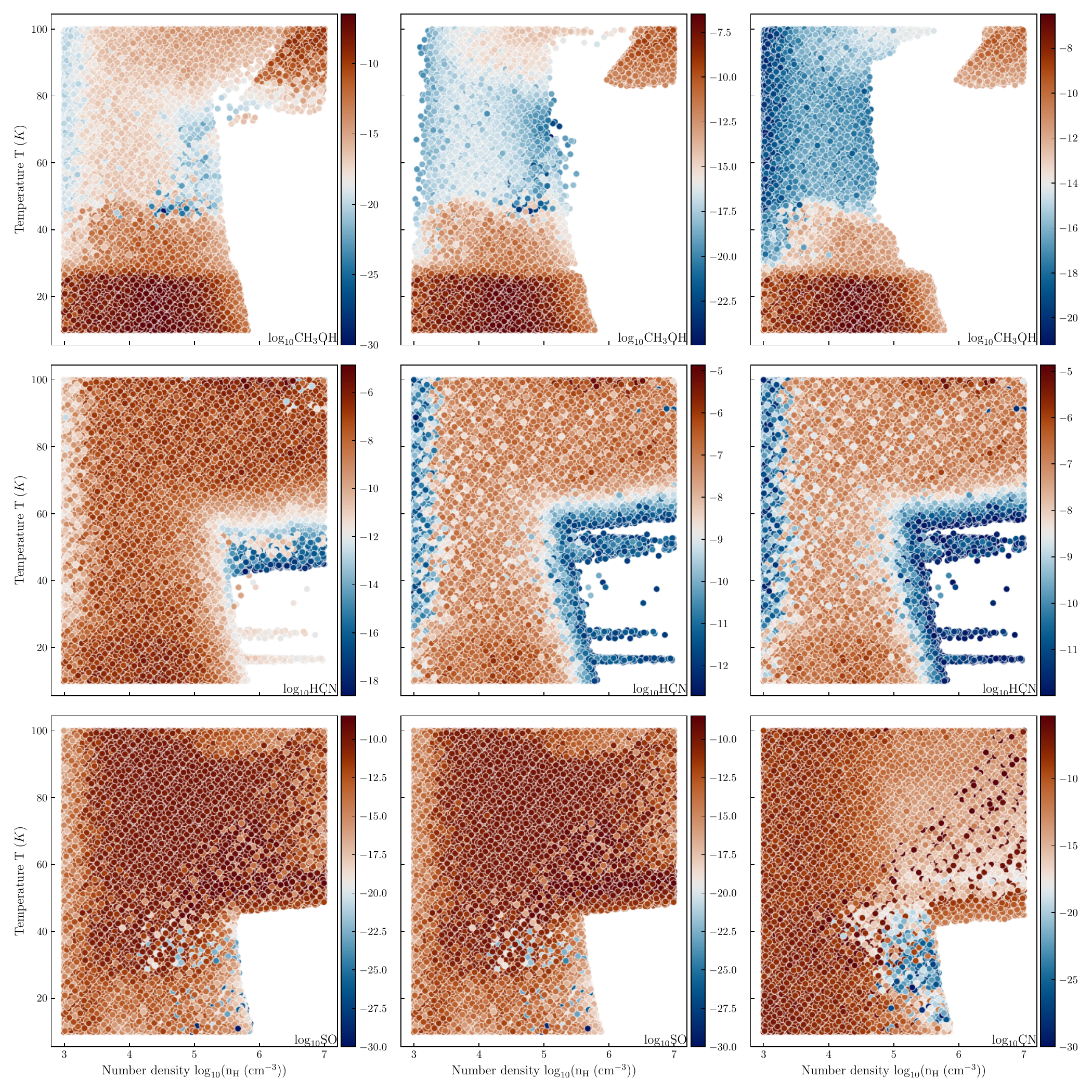}
  \caption{The denominator for the ratios as a function of density and temperature including
  the observational limit.}
  \label{fig:app:denom_nh_t}
\end{figure*}

The individual enumerator and denominator of \Cref{fig:results:ratio-density-temperature} can be found in \cref{fig:app:enum_nh_t} and \Cref{fig:app:denom_nh_t}.
\end{appendix}

\end{document}